\documentclass[lettersize,journal]{IEEEtran}
\usepackage{amsmath,amsfonts}
\usepackage{array}
\usepackage[caption=false,font=normalsize,labelfont=sf,textfont=sf]{subfig}
\usepackage{textcomp}
\usepackage{stfloats}
\usepackage{url}
\usepackage{verbatim}
\usepackage{graphicx}
\usepackage{cite}
\usepackage{diagbox}
\usepackage{color}
\usepackage{multirow}
\usepackage{threeparttable}

\usepackage{amsmath,amssymb,amsfonts}
\usepackage{hyperref}
\usepackage{algorithm}  
\usepackage{algpseudocode}  

\newcommand\etal{\emph{et al.}}
\newcommand{\vct}[1]{\boldsymbol{#1}} 
\newcommand{\mat}[1]{\boldsymbol{#1}} 

\begin{document}

\title{A Dual-Level Cancelable Framework for Palmprint Verification and Hack-Proof Data Storage}


\author{Ziyuan~Yang, Ming~Kang, Andrew~Beng~Jin~Teoh,~\IEEEmembership{Senior Member,~IEEE,} Chengrui Gao, Wen~Chen, Bob~Zhang,~\IEEEmembership{Senior Member,~IEEE,} Yi Zhang,~\IEEEmembership{Senior Member,~IEEE}
        

\thanks{This work was supported in part by the National Natural Science Foundation of China under Grant 62271335; in part by the Sichuan Science and Technology Program under Grant 2021JDJQ0024; in part by the Sichuan University “From 0 to 1” Innovative Research Program under Grant 2022SCUH0016.~\textit{(Corresponding author: Yi Zhang)}}
\thanks{Ziyuan Yang is with the College of Computer Science and Key Laboratory of Data Protection and Intelligent Management, Ministry of Education, Sichuan University, Chengdu 610065, China (e-mail: cziyuanyang@gmail.com)}
\thanks{Ming Kang and Wen Chen are with the School of Cyber Science and Engineering, Sichuan University, Chengdu 610065, China (e-mail: mkang9464@163.com; wenchen@scu.edu.cn)}
\thanks{Andrew Beng Jin Teoh is with the School of Electrical and Electronic Engineering, College of Engineering, Yonsei University, Seoul, Republic of Korea (e-mail: bjteoh@yonsei.ac.kr)}
\thanks{Chengrui Gao is with the College of Computer Science, Sichuan University, Chengdu 610065, China (e-mail: cr@stu.scu.edu.cn)}
\thanks{Bob Zhang is with the Pattern Analysis and Machine Intelligence Group, Department of Computer and Information Science, University of Macau, Taipa, Macau, China (e-mail: bobzhang@um.edu.mo)}
\thanks{Yi Zhang is with the School of Cyber Science and Engineering and Key Laboratory of Data Protection and Intelligent Management, Ministry of Education, Sichuan University, Chengdu 610065, China (e-mail: yzhang@scu.edu.cn)}}

\markboth{Journal of \LaTeX\ Class Files,~Vol.~14, No.~8, August~2021}%
{Shell \MakeLowercase{\textit{et al.}}: A Sample Article Using IEEEtran.cls for IEEE Journals}



\maketitle

\begin{abstract}
In recent years, palmprints have been widely used for individual verification. The rich privacy information in palmprint data necessitates its protection to ensure security and privacy without sacrificing system performance. Existing systems often use cancelable technologies to protect templates, but these technologies ignore the potential risk of data leakage. Upon breaching the system and gaining access to the stored database, a hacker could easily manipulate the stored templates, compromising the security of the verification system.
To address this issue, we propose a dual-level cancelable palmprint verification framework in this paper.
Specifically, the raw template is initially encrypted using a competition hashing network with a first-level token, facilitating the end-to-end generation of cancelable templates. Different from previous works, the protected template undergoes further encryption to differentiate the second-level protected template from the first-level one.
The system specifically creates a negative database (NDB) with the second-level token for dual-level protection during the enrollment stage. Reversing the NDB is NP-hard and a fine-grained algorithm for NDB generation is introduced to manage the noise and specified bits. During the verification stage, we propose an NDB matching algorithm based on matrix operation to accelerate the matching process of previous NDB methods caused by dictionary-based matching rules. This approach circumvents the need to store templates identical to those utilized for verification, reducing the risk of potential data leakage.
Extensive experiments conducted on public palmprint datasets have confirmed the effectiveness and generality of the proposed framework. Upon acceptance of the paper, the code will be accessible at \url{https://github.com/Deep-Imaging-Group/NPR}.
\end{abstract}


\section{Introduction}
\label{sec:int}
\IEEEPARstart{R}{ecently}, biometric technology has been ubiquitous for identity management~\cite{jain2021biometrics}. Among various biometrics, palmprint has become prevalent due to its exceptionally rich feature set, universality, and low intrusiveness, making it a popular choice in the field~\cite{Wang2024TSMCS}.

Early studies of palmprint verification focused on manually designing texture feature extractors and coding rules~\cite{fei2016palmprint, yang20212TCC,fei2016half}. However, these methods highly relied on prior knowledge and often yielded poor accuracy. Inspired by the success of deep learning (DL), researchers have proposed several DL-based palmprint verification methods, which have achieved impressive accuracy~\cite{fei2018feature, dong2022co,su2023learning}.

Nevertheless, biometric features, commonly known as templates, being inherently immutable and uniquely linked to an individual, pose major security and privacy challenges if exposed. Therefore, it is necessary to adopt biometric template protection (BTP) techniques to mitigate this vulnerability. Cancelable biometrics (CB) is one such technique~\cite{lee2022alignment}. CB employs an irreversible transformation with a user-specific token to create protected biometric templates. This method allows users to revoke and regenerate new templates by altering the user-specific token if compromised.


As per the ISO/IEC 30136 standard~\cite{isoiec30136}, the BTP scheme, including CB, must meet the following criteria:
\begin{itemize}
    \item \textit{Cancelability}. The scheme should allow invalidation and replacement of a compromised biometric template with a new one.
    \item \textit{Unlinkability}. Multiple distinct protected templates should be creatable from a single identity, preventing correlation to the same identity. This allows for template renewal and use across different applications without cross-matching.
    \item \textit{Irreversibility}. Reversing the original biometric templates from their protected forms should be computationally difficult or nearly impossible.
    \item \textit{Accuracy Performance}. The integration of BTP into a biometric verification system must not significantly reduce its verification accuracy.
\end{itemize}

\begin{figure*}
  \centering
  \begin{minipage}[t]{0.47\linewidth}
  \centering
  \includegraphics[width=\textwidth]{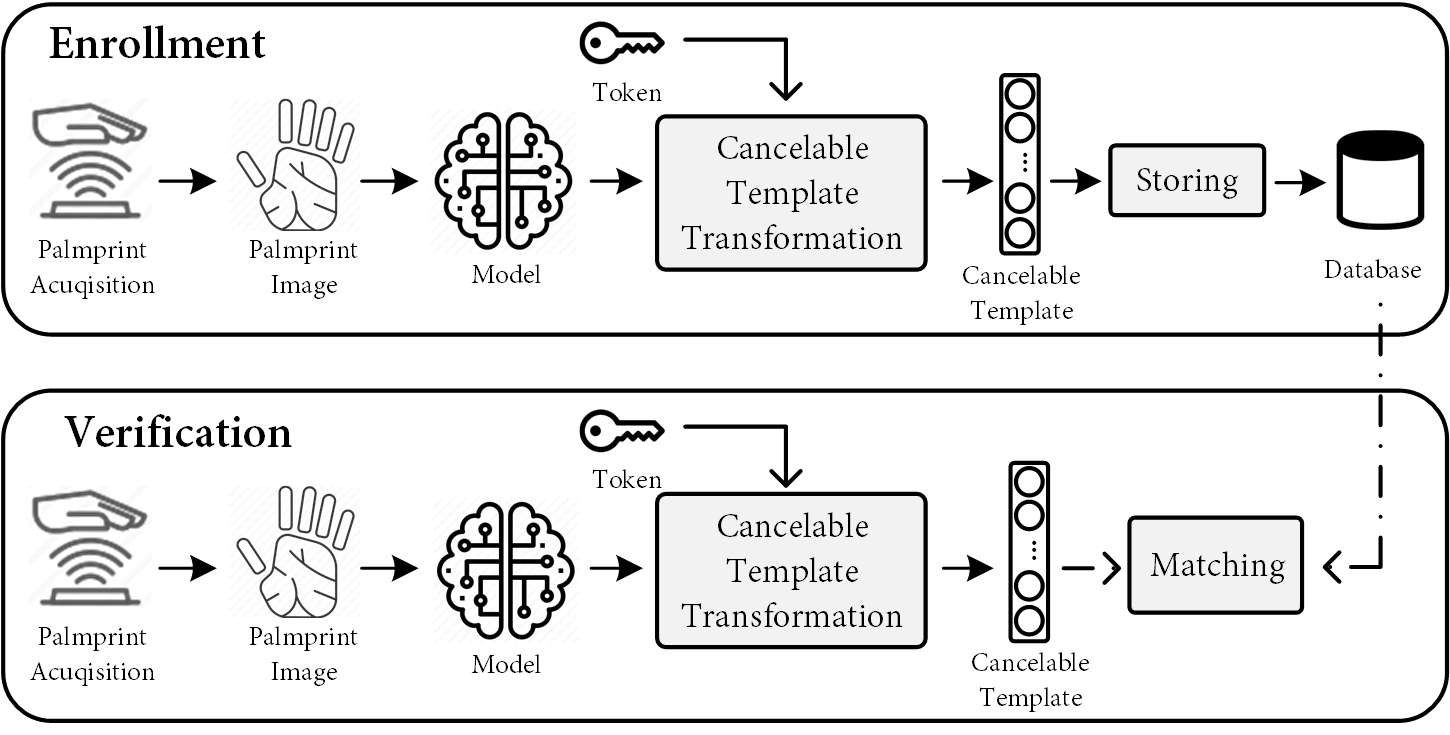}
   \centerline{(a)}
   \label{fig:old_frame}
   \end{minipage}  
   \quad 
  \begin{minipage}[t]{0.47\linewidth}
  \centering
  \includegraphics[width=\textwidth]{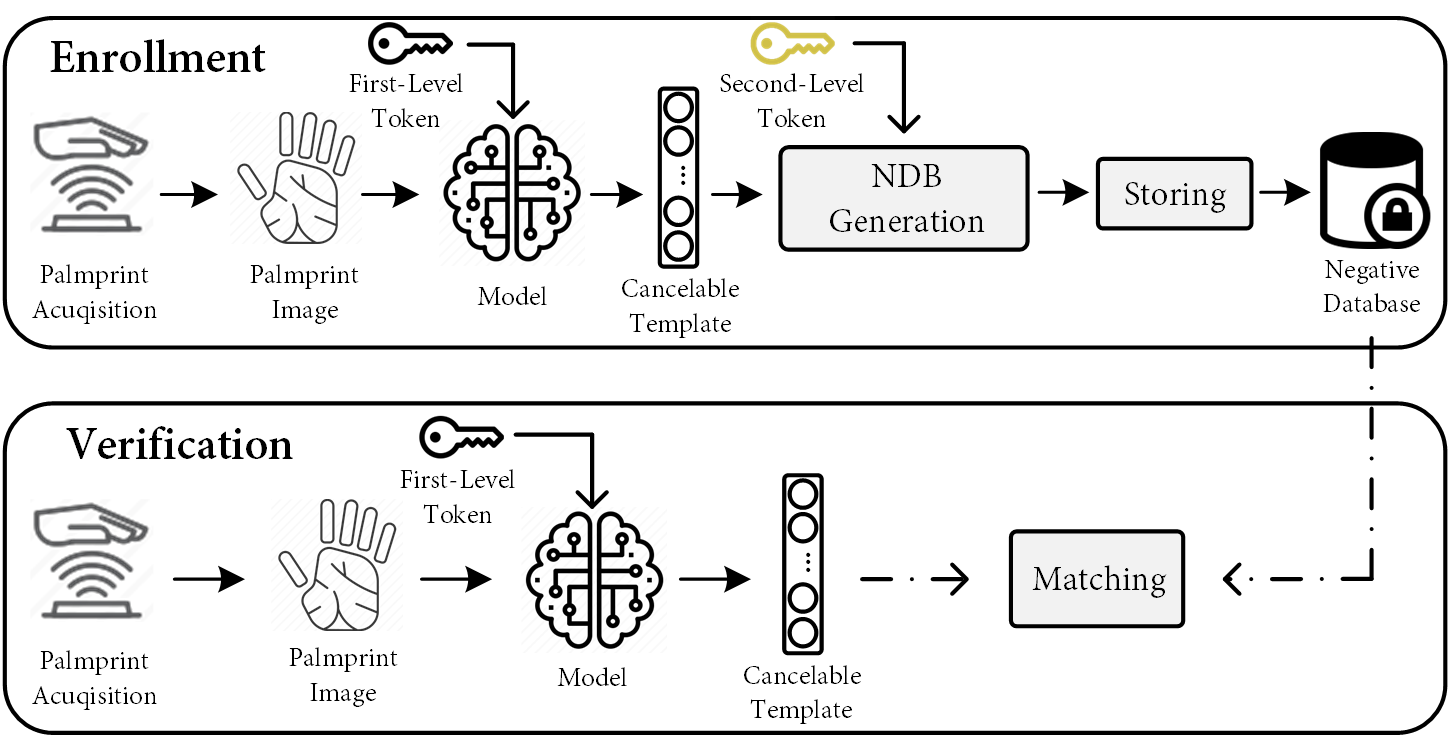}
   \centerline{(b)}
   \label{fig:our_frame}
   \end{minipage} 
     \caption{(a) Traditional Cancellable Palmprint Verification System. (b) Our Proposed Dual-Level Cancelable Palmprint Verification System.}
     \label{fig:sys}
  \label{fig:Concept}
\end{figure*}

\begin{figure}
  \centering
  \includegraphics[width=.9\columnwidth]{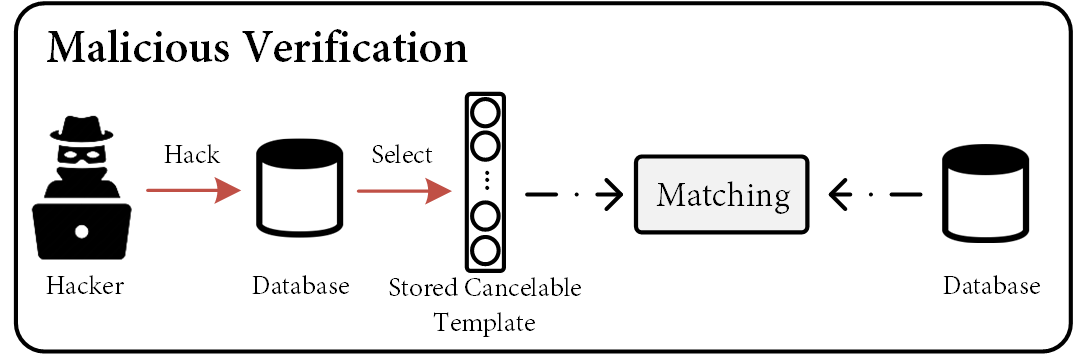}
     \caption{The attack pipeline of the malicious verification}
  \label{fig:malicious}
\end{figure}

Despite the implementation of BTP techniques, a significant security concern at the template storage level has been consistently overlooked. Especially during the enrollment phase, it is crucial for the system to store the template in the same format as the query template for matching, as depicted in Fig.~\ref{fig:sys}(a). If an adversary gains access to the stored database, they could easily use the stored template corresponding to a target ID as a query template for matching. Such unauthorized access seriously threatens the integrity of the biometric system, compromising its security. To illustrate such an attack, we visually depict the malicious verification process in Fig.~\ref{fig:malicious}.



To address the challenge mentioned above, this paper introduces a novel \textbf{\textit{D}}ual-Level \textbf{\textit{C}}ancelable \textbf{\textit{P}}almprint \textbf{\textit{V}}erification Framework, dubbed DCPV. In addition to its primary goal of template protection, DCPV  also focuses on enhancing security with an additional objective - Secure Data Storage.


We introduce a novel Hashing Competition Palmprint Network designed for CB template generation to establish the first layer of protection. This network consists of a competition feature extraction backbone and a template transformation head. Leveraging the effectiveness of competition mechanisms in palmprint verification~\cite{yang2023comprehensive}, we embed a competition mechanism in the network for feature extraction to obtain discriminative features. These features are further extracted and transformed via a user-specific, randomly-weighted multi-layer perceptron.

The resulting protected templates are converted into binary features to minimize storage overhead. However, directly converting real-number features into binary may reduce accuracy performance. To overcome this challenge, we employ a contrastive loss, which encourages the network to produce discriminative binary features, balancing accuracy with feature dimensionality.

To mitigate the security risks associated with the direct storage of CB templates, we propose a secondary safeguard by implementing a Negative Database (NDB) notion~\cite{zhao2015fine}. Rather than directly storing the actual CB templates, we use a second-level token to store their negative (complementary) counterparts. This entails establishing an NDB containing negative codes derived from the CB templates, rendering them infeasible from being directly used as query templates for verification.

Despite the potential risk of unauthorized access by adversaries to the stored database, reversing an NDB is an NP-hard problem, offering a solid theoretical foundation for robust privacy and security measures against such attacks~\cite{zhao2021k}. However, traditional retrieval from the NDB, based on time-consuming dictionary rules for matching, significantly lowers verification efficiency. To overcome this limitation, we propose implementing a matrix-computing-based matching rule to improve the speed and efficiency of the verification process.

The proposed DCPV distinguishes itself from previous cancelable palmprint verification methods, as exemplified in Fig.~\ref{fig:sys}(b). A key feature of DCPV is ensuring a noticeable format difference between the stored template and that used during user verification. Unlike previous approaches~\cite{leng2015alignment, leng2011dual}, the double-protected templates are not suitable for direct verification. Furthermore, the non-reversibility of NBD further enhances the non-irreversibility of the original templates. This approach offers a versatile solution for enhanced template protection, enabling modifications at both the first and second levels of tokens. This adaptability renders DCPV effective in addressing a variety of security-sensitive challenges.



The main contributions of this paper can be summarized as follows:
\begin{itemize}
    \item We propose a novel dual-level cancelable palmprint verification framework for fine-grained protection.
 
    \item We introduce the concept of NDB into palmprint verification to relieve the threat of dataset leakage to system security.

    \item Extensive experiments are conducted to assess the effectiveness of the proposed method across multiple public palmprint verification datasets.
\end{itemize}

The rest of this paper is organized as follows: Sec.~\ref{sec:rela} presents an overview of related works. The detailed methodology is elaborated in Sec.~\ref{sec:meth}. Comprehensive experiments and a discussion of their outcomes are given in Sec.~\ref{sec:exp}. Finally, the paper is concluded in Sec.~\ref{sec:con}.

\section{Related Works}
\label{sec:rela}
\vspace{5pt}
\subsection{Palmprint Verification}
Over the past decades, there has been a prolific influx of palmprint verification research~\cite{fei2023learning}. Encouraged by the potent texture extraction capabilities of Gabor filters, researchers have proposed enormous coding-based methods~\cite{ZhongTIFS}. Typically, Zhang~\etal~\cite{zhang2003online} used a single Gabor filter along 45$^{\circ}$ to extract magnitude features, which were subsequently encoded to yield binary templates. Additionally, many methods based on magnitude features have been proposed~\cite{guo2009palmprint,zhang2012fragile,kong2006palmprint,yang2021extreme,yang20212TCC}. However, these methods are sensitive to illumination change~\cite{fei2020feature}. 

A competition mechanism has been introduced into palmprint verification to mitigate illumination change issues and achieved remarkable success~\cite{kong2004competitive}. Competition mechanism-based methods mainly extract and code the index of the winner orientation as the feature~\cite{jia2008palmprint,fei2016half,fei2016double,xu2018drcc}. However, these early works heavily relied on the prior knowledge of the researchers, which may hinder the performance.

To relieve the high reliance on prior knowledge, researchers have proposed DL-based palmprint verification methods~\cite{shao2022towards, zhao2020joint}. For example, Genovese~\etal~\cite{genovese2019palmnet} proposed an unsupervised palmprint network known as PalmNet, which contains multiple feature extraction branches, including convolutional neural network (CNN), Gabor filters, and principal component analysis. Zhong and Zhu~\cite{zhong2019centralized} proposed centralized large-margin cosine loss to optimize the distribution of samples in the feature space. Zhang~\etal~\cite{zhong2018palm} and Wu~\etal~\cite{wu2021palmprint} developed deep hashing network (DHN) in palmprint verification. Jia~\etal~\cite{jia2022eepnet} introduced a lightweight palmprint network, EEPNet. Liang~\etal proposed a novel network~\cite{liang2023pklnet}, aiming to attain superior verification performance by integrating information from various regions of the hands. Besides, encouraged by the success of the competition mechanism, researchers proposed competition palmprint verification networks and achieved satisfactory performance~\cite{yang2023co3net,liang2021compnet}. For example, Yang~\etal~\cite{yang2023comprehensive} extended the competition mechanism to extract comprehensive competition features, including channel, spatial, and multi-order competitive features. To relieve the data privacy concerns in training, the authors proposed to combine federated learning into palmprint recognition~\cite{shao2023privacy}. Besides, inspired by the physical characteristics, authors proposed to achieve spectrum consistency in the federated training process~\cite{yang2023physics}. However, these methods lack protection for the extracted features~\cite{yang2023cross}, which may raise security and privacy concerns.

\begin{figure*}
\vspace{5pt}
  \centering
  \includegraphics[width=\textwidth]{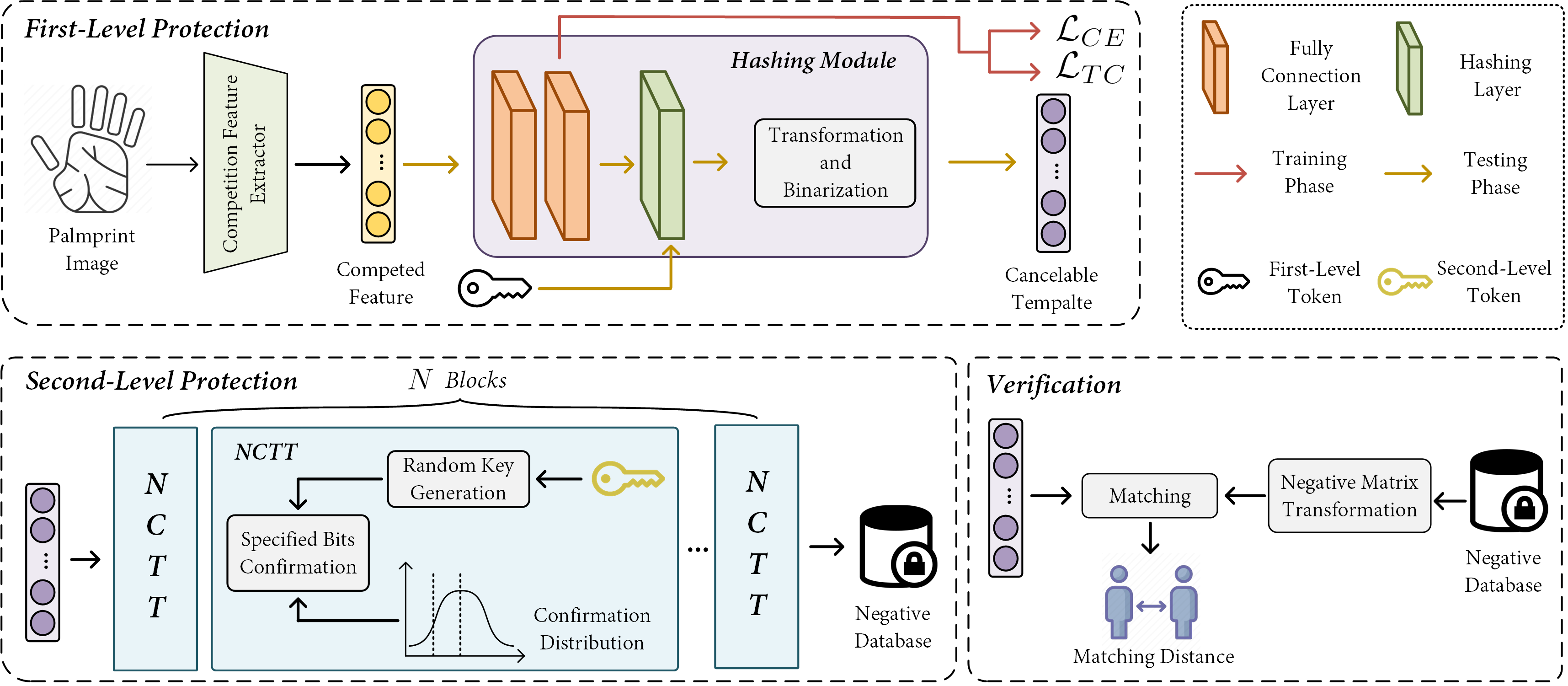}
   \label{fig:Concept}
     \caption{The overview of the proposed methods. The proposed DCPV is a dual-level protection framework. In the first-level protection, the model trained with a hybrid loss would generate a binarized cancelable template with the first-level token. Then, the system would store the set of the negative form of the first-level template. In this way, for the matching pairs, the query and stored templates are different in the verification stage. Hence, DCPV could alleviate the potential risk of unauthorized access by adversaries to the stored database.}
  \label{fig:Overview}
\end{figure*}



\subsection{Cancelable Biometrics for Palmprint}

In the expansive realm of CB, this subsection centers specifically on advances in cancellable biometric techniques tailored for palmprint verification. It is crucial to note that while this review delves deeply into CB specific to palmprint, the literature on CB spans various modalities. For a comprehensive overview of CB across different biometric types, readers are recommended to refer the comprehensive survey papers, such as~\cite{bernal2023review} and~\cite{manisha2020cancelable}. 


Qiu~\etal~\cite{qiu2019cancelable} employed an anisotropic filter to capture the orientation feature, which is then measured by a chaotic matrix to generate the cancelable palmprint template.
Ashiba~\etal~\cite{ashiba2023proposed} utilized a homomorphic filtering masking encoding algorithm to protect palmprint templates. Besides, Lee~\etal~\cite{lee2024extraction} proposed to transform the intersection points of palmprint and palm-vein for template protection. A novel randomized cuckoo hashing is applied on the binary palmprint feature to enhance security~\cite{li2020palmprint}. Salt-hashing (SH)-based CB methods can protect the token template independent of the biometrics. Biohashing~\cite{jin2004biohashing} was the first attempt to introduce SH-based technology into biometrics, generating a user-specific random matrix to randomly project the template into a new feature space. Sujitha and Chitra~\cite{sujitha2019novel} proposed a fingerprint and palmprint-based multi-biometric cryptosystem using texture features extracted via bottom-hat filters. Besides, Jin~\etal~\cite{jin2017ranking} proposed an Index-of-Max hashing method to project the features using the indices of ranked features. 


In recent years, DL-based CB methods have also gained significant attention. For instance, deep table-based hashing~\cite{jang2019deep} is proposed to generate binarized hashing codes from biometric data. However, this approach is impractical and insecure~\cite{dong2022deep}. Subsequently, Dong~\etal~\cite{dong2022deep} proposed a deep-rank hashing network to address these issues in the identification task. Besides, Wu~\etal~\cite{wu2023multi} embed the fuzzy commitment technology into the deep hashing network to protect the palmprint templates, although this method necessitates multiple spectrum templates. Shahreza~\etal~\cite{shahreza2023mlp} suggested using a hashing multi-layer perceptron (MLP) for projecting templates into a new space. 

Zhao~\etal~\cite{zhao2015negative} first introduced the NDB technology, $p$-hidden~\cite{liu2014p}, into the CB field and achieved satisfactory results. However, these methods require storing the verification template in the system, raising potential privacy concerns.

\section{Methodology}
\label{sec:meth}
\vspace{5pt}
\subsection{Overview of the Proposed DCPV}



As previously mentioned, current CB schemes may not effectively protect against attacks targeting the stored database. To address this vulnerability, DCPV is specifically crafted to address data storage concerns through a dual-level protection approach, which integrates CB transformation and NDB notion, as illustrated in Fig.~\ref{fig:Overview}.




\subsection{Hashing Competition Palmprint Network}



We introduce a trainable competition palmprint feature extractor. This extractor utilizes learnable Gabor filters to capture distinctive texture features and integrates a comprehensive competition mechanism to generate resilient competed features~\cite{yang2023comprehensive}. It is crucial to highlight that the features extracted through this process lack protection. Consequently, our effort focuses on transforming the unprotected feature into a cancelable instance by introducing a hashing module, thereby achieving first-level protection.



 Let $\vct{u}\in \mathbb{R}^{m_f}$ be the unprotected feature vector extracted by the competition feature extractor, where $m_f$ denotes the vector's length. 
 Additionally, $k_1$ represents the first-level token.
 


Subsequently, we integrate the hashing layer into the competition network, positioned after the final standard linear layer. Specifically, for the final linear layer, we initiate the process by generating a pseudo-random matrix $\mat{M}_f$ based on $k_1$, where $\mat{M}_f\in\mathbb{R}^{m_f\times m_p}$ and $m_p$ is the length of the protected template. The Gram-Schmidt process~\cite{jin2004biohashing} is then applied to the rows of $\mat{M}_f$, transforming it into an orthonormal matrix denoted as $\mat{M}_{\perp}$. Then, the protected code is obtained through the inner product operation:
\begin{equation}
    \vct{q}_j = \langle \vct{u}, \mat{M}_{\perp} \rangle,
\end{equation}
where $\vct{q}_j$ represents the random transformed $\vct{u}_j$, and $\vct{q}_j \in \mathbb{R}^{m_p}$.

Once the first-level protected feature $\vct{q}$ is acquired, it undergoes binarization to enhance security and mitigate the storage overhead. Generally, the process can be formulated as follows:

\begin{equation}
    \vct{b}(h)=\left\{\begin{array}{lll}
0 & \text { if } & \mat{q}(h) \leq \beta \\
1 & \text { if } \quad & \mat{q}(h)>\beta
\end{array}\right.,
\end{equation}
where $\beta$ denotes the median value of $\vct{q}$. $\vct{b}\in \mathbb{R}^{m_p}$ denotes the binarized protected template and $h$ denotes the index of the feature. The protection discussion of this way can be found in~\cite{shahreza2023mlp}.

Nevertheless, feature binarization poses a potential issue to the discriminative capacity of features, especially when opting for naive binarization as in Eq. (2), which might impede network convergence during training. To overcome this challenge, we combine the $tanh(\cdot)$ activation function with the supervised contrastive loss~\cite{khosla2020supervised}. This approach is implemented to preserve the discriminative capability of features throughout the training process. As a result, the features are guided toward values of 0 or 1, ensuring that the binarized features do not excessively compromise their inherent discriminative capacities. The supervised contrastive loss is given as follows:
\begin{equation}
    \mathcal{L}_{TC} = -\sum_{i \in I} \frac{1}{|K(i)|} \sum_{p \in P(i)} \log \frac{\exp \left(\vct{z}_{i} \cdot \vct{z}_{p} / \tau\right)}{\sum_{a \in A(i)} \exp \left(\vct{z}_{i} \cdot \vct{z}_{a} / \tau\right)},
    \label{eq:ltc}
\end{equation}
where $I\equiv\{1, \ldots, 2 N_{bs}\}$ is the batch of contrastive sample pairs, $A(i) \equiv I \backslash\{i\}$ and $i$ is the index of positive sample. $P(i)\equiv {p \in A(i): y_i = y_p}$ is the index set of the positive samples in the batch distinct from $i$, and $y_i$ is the label of the $i$-th sample in the batch. $|K(i)|$ denotes the number of samples in $K(i)$. $\vct{z}_{i}$ and $\vct{z}_{p}$ stand for the anchor and positive features, respectively. $\tau$ is the temperature parameter. Furthermore, $\vct{z}=tanh(\mat{Q})$. Besides, the cross-entropy loss is used as the task loss, which can be formulated as:
\begin{equation}
    \mathcal{L}_{CE} = -\frac{1}{T}\sum_{i=1}^T\sum_{c=1}^{S} y_{i,c} \log \left(q_{i,c}\right),
    \label{eq:ce}
\end{equation}
where $T$ and $S$ denote the numbers of samples and classes, respectively. $y_{i,c}$ and $q_{i,c}$ represent the label and the predicted probability of the $i$-th sample, respectively. Finally, our proposed loss function is given as follows:
\begin{equation}
    \mathcal{L}= w \times \mathcal{L}_{CE} + (1-w)\times \mathcal{L}_{TC},
    \label{hy_loss}
\end{equation}
where $w$ is the weight of $\mathcal{L}_{CE}$. Empirically, in this paper, $w$ is set to 0.8.

This strategy allows the model to extract discriminative features that preserve their distinctiveness even post-binarization. As a result, this hybrid loss effectively alleviates the performance decline typically associated with cancelable mapping in the network.




  \begin{algorithm}[t]  
  \caption{Negative Cancellable Template Transformation Algorithm.} 
  \label{alg:Framework}  
  \begin{algorithmic}[1]  
\Require A $m$-bit cancelable feature $\vct{b}$, the second-level token $k_2$, the length of the generated NDB $N$, and the confirmation distribution $P$.
    \Ensure Real-valued negative cancellable template $\mat{B_r}^g$.
    \State \textbf{Enrollment Function:} 
    \State Receive $\vct{b}$.
    \State $\mat{B}^g \gets \emptyset$
    \State Set the seed as $k_2$.
    \For{$z\leq N$}
    \State $\vct{b}^g_z \gets$ \textbf{NCTT}($\vct{b},P$)  
    \State $\mat{B}^g \gets \mat{B}^g \cup \vct{b}^g_z$
    \EndFor
    \State Convert string-based $\mat{B}^g$ to real-valued-based $\mat{B}_r^g$.\\
    \Return $\mat{B}_r^g$
    \State \textbf{Function NCTT}($\vct{b}, P$)
    \State Generate the random value $v_r$ and $0 < v_r < 1$.
    \State Obtain the index $i$ of $v_r$ within the $i$-th interval within $P$.
    \State Initlize $\vct{b}^g$ as a $m$-length string, consisting entirely of $*$.
    \State Select $i$ indices from $\vct{b}^g$ such that values at these indices differ from those in $\vct{b}$.
    \State Select $K-i$ indices from $\vct{b}^g$ such that values at these indices are the same as those in $\vct{b}$.\\
    \Return $\vct{b}^g$
  \end{algorithmic}  
\end{algorithm}

\subsection{Negative Cancellable Palmprint Templates}


While the hashing competition palmprint network is employed for template protection, its limitations, especially regarding database leaks, prompt us to avoid directly storing biometric templates. Instead, our approach stores the negative counterpart of the first-level CB template using a second-level token $k_2$. The negative dataset is well-established as an NP-hard problem, providing a robust theoretical foundation for privacy preservation~\cite{esponda2007protecting,esponda2004online,jiang2023negative}. In this paper, we employ the $K$-hidden method for negative database generation to create a negative CB template~\cite{zhao2015fine}.

\subsubsection{\textbf{Negative Templates Transformation}}
The CB template $\vct{b} \in \{0, 1\}^m$ is transformed into negative CB template, denoted as $\mat{B}^g \in \mathbb{R}^{N \times m}$. $\mat{B}^g$ is composed of $N$ negative strings, $\vct{b}^g \in \mathbb{R}^{m \times 1}$, where $N=m\times r$ and $r$ is a hyperparameter. $\vct{b}^g$ is composed of types of characters, which are ``0", ``1" and ``*", respectively. For fine-grained control over the probabilities of generating $K$ distinct types of negative string, we randomly initialize an interval set denoted as $P$, which consists of $K$ intervals,
with the $i$-th interval expressed as $[p_{i-1},p_i)$, and $p_1+...+p_K=1$.


To execute negative template transformation, each \(\vct{b}^g \in \Sigma^m\) initializes a string consisting entirely of ``*" to symbolize an unspecified character. Subsequently, a random value \(v_r\) is generated with the second-level token \(k_2\), ensuring \(0<v_r<1\). Each \(\vct{b}^g\) is then constructed with \(K\) specified characters, and the indices of these specified characters are also selected using the second-level token \(k_2\). The index \(i\) is identified within the \(i\)-th interval of \(P\). Once \(i\) is determined, the structure of \(\vct{b}^g\) is established: it comprises \(i\) specified characters differing in value from those in \(\vct{b}\) and \(K-i\) specified characters matching the values in \(\vct{b}\). This process completes the generation of \(\vct{b}^g\), and the system repeats these operations \(N\) times to generate the negative cancellable template \(\mat{B}^g\), where \(\mat{B}^g = [\vct{b}_1^g,...,\vct{b}_N^g]\).

\begin{table}[]
    \centering
        \caption{The Dictionary-based Matching Rules of the Previous NDB Methods.}
        \small
    \begin{tabular}{rcccccc}
    \hline
       Negative Bit   & 0 & 1 & 0 & 1 & * & *\\
     Query Bit      & 0 & 1 & 1 & 0 & 0  & 1\\ 
     Matching Distance      & -1 & -1 & 1 & 1 & 0 & 0 \\ \hline
    \end{tabular}
    \label{Neg_Rule}
\end{table}

\subsubsection{\textbf{String Conversion and Matching}}

The above operations necessitate a transition from real-number-based to string-based codes. As a result, as described in~\cite{zhao2015fine}, the matching relies on dictionary-based matching rules to calculate the distance between the query CB code and the negative CB template. 
We refer to the bit in the negative code as the ``Negative Bit" and the bit in the query code as the ``Query Bit". The matching distances for various pairs of Negative Bit and Query Bit are detailed in Tab.~\ref{Neg_Rule}. The cumulative sum of the matching distances is considered the final matching result after completing the matching process between the query code and all negative codes. Unfortunately, this verification process is time-consuming.

To address this challenge, we propose an expedited matrix-computing-based matching approach. Specifically, we transform the string-based matrix \(\mathbf{B}^g\) into a real-valued matrix \(\mathbf{B}_r^g \in \mathbb{R}^{N \times m}\) by converting ``*" and ``0" into the real number 0 and -1, respectively, while directly transforming ``1" into the real number 1. For example, the transformation of a negative code ``01*" results in the matrix \([-1, 1, 0]\). Using this approach, the correlation between the matching results derived from our proposed matching rules and those obtained through dictionary-based matching is characterized by their values being opposite numbers.

As an illustration, the initial matching scores for string pairs ``1" with ``1" and ``0" with ``0" are -1. However, these scores become 1 after our string-to-real-valued conversion, as shown by the equations \(1 = -1 \times -1\) and \(1 = 1 \times 1\). Therefore, equivalent matching results are achieved by simply reversing the matching score.




In this paper, the matching score $d$ between the query CB code $\vct{b}_q\in \mathbb{R}^{m\times 1}$ and $\mat{B}_r^g$ is calculated as follows:
\begin{equation}
    d = arccos(-sum(\mat{B}_r^g \cdot \vct{b}_q)),
\end{equation}
where $sum(\cdot)$ represents the cumulative total of all elements in the input.

This matching process, involving a matrix and a vector, presents a non-trivial problem due to the inherent dimensional disparity between the two. Meanwhile, the non-trivial challenge extends to the transformation between the generated negative form and its original form, adding another layer of sophistication to the complex matching scenario. To help the readers understand the negative cancellable template transformation, the pseudocode is provided in Alg.~\ref{alg:Framework}.

\subsection{Security Analysis}
In this subsection, we will give a security analysis of the negative database, which is crucial to ensure irreversibility. 

\textbf {Lemma 1.} Given a generated NDB $NDB_s$, and a random $m$-bit string $y_s$ that has $u$ bits different from the original feature string $s$, the expected proportion of entries in $NDB_s$ that do not match $y$ can be formulated as:
\begin{equation}
    g(\alpha)=1-\sum_{i=1}^{K}p_{i}\times\alpha^{i}\times(1-\alpha)^{K-i}.
\end{equation}
 
 \indent \textit {Proof.} The probability that $NC$ matches $y_s$ is: 
 
 \begin{equation}
    C_\mathrm{iy}=\frac{\binom{m-K}{u-i}}{\binom mu}\xrightarrow{m\to\infty}\frac{u}{m}^i(1-\frac{u}{m})^{K-i}.  
 \end{equation}\par

 For simplicity, we define $\alpha = \frac{u}{m}$. The NDB algorithm's probability of type $i$ is $p_i$. Then, the expected proportion of entries in N DBs that do not match $y_s$ is:
 
 
 
 \begin{equation}
    g(\alpha)=1-\sum_{i=1}^{K}p_{i}\times\alpha^{i}\times(1-\alpha)^{K-i}.
 \end{equation}\par
 
 Then, we can get the derivative of $g(\alpha)$ through Lemma 1 as:
 \begin{equation}
    g^{\prime}(\alpha)=\sum_{i=1}^K(K\alpha-i)\times p_i\times\alpha^{i-1}\times(1-\alpha)^{K-i-1}.
 \label{e10}
 \end{equation}\par
 
The minimal positive solution $\alpha_0$ of $g^{\prime}(\alpha)=0$ is the smallest positive stationary point of $g(\alpha)$.\par

\textbf {Lemma 2.} Assume $P_s$ is the probability that the local search strategy successfully finds the string $s$ hidden by $NDB_s$, then we have:

\begin{equation}
    P_s\xrightarrow{m\to\infty}
    \begin{cases}0&\alpha_0<\frac{1}{2}\\\frac{1}{2}&\alpha_0=\frac{1}{2}\\1&\alpha_0>\frac{1}{2}
    \end{cases}.
\end{equation}

\textit{Proof.} The details can be found in~\cite{liu2014p}.

\textbf {Theorem 1.} The $K$-hidden-NDB is hard to reverse with regard to the local search strategy when the following condition is satisfied:

\begin{equation}
    \sum_{i=1}^{K}(K-2i)p_{i}>0.
\label{e12}
\end{equation}

\indent \textit {Proof.} According to Eq.~\eqref{e10}, when $\alpha{<}1/K$, we have $(K\alpha-i)<0$.
When $\alpha = 1 / k$, we have:

 \begin{equation}
    g'\biggl(\frac{1}{K}\biggr)=\sum_{i=1}^{K}(1-i) p_{i}\biggl(\frac{1}{K}\biggr)^{i-1}\biggl(1-\frac{1}{K}\biggr)^{X-i-1}\le0.
\end{equation}

When $\alpha=1/2$ and $\sum_{i=1}^{K}(K-2i)p_i>0$, we have:

\begin{equation}
    g^{\prime}\biggl(\frac{1}{2}\biggr)=\biggl(\frac{1}{2}\biggr)^{K-1}\sum_{i=1}^{K}\left(K-2i\right)p_{i}>0.
\end{equation}\par

Consequently, we can ascertain that at least one value of $\alpha$ exists within the interval $[1/K,1/2]$ (given that $(K\geq3)$) such that $g^{\prime}(\alpha_{1})=0$. This implies that the function $g(\alpha)$ has at least one positive stationary point less than $1/2$. Therefore, under the condition that Eq.~\ref{e12} holds true, we deduce that $\alpha{_0}<1/2$. This leads to the generated $NDB_s$ being challenging to reverse with respect to the local search strategy in alignment with Lemma 2. It is important to note that while Eq.~\ref{e12} represents a sufficient condition for this scenario, it is not necessary when $K>3$; however, in the case where $K=3$, Eq.~\ref{e12} serves as both a sufficient and necessary condition, paralleling the hardness condition outlined in~\cite{liu2014p}.



\subsection{Implementation}

\noindent The proposed hashing competition palmprint network is initially trained using a hybrid loss function, as given in Eq.~\eqref{hy_loss}. After training, the model performs feature extraction and CB transformation during enrollment, creating a first-level CB template with a corresponding token. To enhance database security against potential hacking threats, an additional layer of protection conceals the CB template. Then, user verification is accomplished by matching the first-level CB template with the stored second-level protected cancelable template.

Notably, the second-level protected template in the system consists of negative codes generated from the first-level CB template. The stored negative codes are significantly different from the first-level CB template, making them ineffective for individual verification purposes. Additionally, reversing a NDB is an NP-hard problem. Even if an adversary accesses the negative codes of a targeted template, reversing them to retrieve the original template is extremely challenging. This dual-level security strategy safeguards against unauthorized access and reverse attack attempts.

Furthermore, the DCPV features a dual-level cancelable framework. This means that DCPV can offer fine-grained protection for biometric templates, suitable for diverse scenarios, like cloud-based verification. Specifically, users can manage the first-level protection, while the server handles the second-level protection. This deployment enables the server to periodically update stored templates without user intervention, such as key alterations. As a result, this framework enhances the resilience of the verification system against potential database breaches, thus establishing a more robust defense against unauthorized access.

\section{Experimetns}
\label{sec:exp}
\subsection{Experimental Setting}
\vspace{5pt}
\noindent To validate the proposed DCPV framework, we conducted experiments on several public palmprint verification datasets, including PolyU~\cite{zhang2003online}, Multi-Spectral~\cite{zhang2009online}, and IITD~\cite{kumar2010personal}. The Multi-Spectral dataset comprises four sub-datasets: red, green, blue, and near-infrared red (NIR) spectral datasets. Besides, PolyU and IITD are contact-based and contactless-based palmprint datasets, respectively. By evaluating the proposed DCPV on these different types of datasets, we aimed to assess its effectiveness and applicability comprehensively. The following are the brief description for each dataset:
\begin{itemize}
    \item PolyU~\cite{zhang2003online}: This dataset
contains 7752 images of 386 palms from 193 individuals. These images were collected in two sessions. There are a total of 37,800 genuine and 14,250,600 imposter matching times. 

    \item Multi-Spectral~\cite{zhang2009online}: There are four subdatasets in Multi-Spectral as introduced earlier, and these subdatasets were collected by four different spectrum devices. Each subdatasets contains 600 palms, and there are 12 images collected from each palm. The total genuine and imposter matching times in each subdataset are 18,000 and 8,982,000.

    \item IITD~\cite{kumar2010personal}: In IITD, there are around 5 images acquired from each palm and 2,601 images in total. The number of total matchings is 1,269,600, in which there are 2,760 genuine matchings and 1,266,840 imposter matchings.
\end{itemize}

The proposed DCPV was implemented using PyTorch. For IITD and PolyU, the number of learning epochs was set to 3000, while for the Red, Green, Blue, and NIR sub-datasets, it was set to 1000. We utilized the Adam optimizer~\cite{kingma2014adam} with a learning rate of 0.01. For all datasets, the 1st session data were used for training; while the 2nd session data were used for testing. The hardware used for this research included an AMD Ryzen 7 5800X CPU @3.80 GHz, 32 GB of internal storage, and an NVIDIA GTX 3080 Ti GPU.

\subsection{Verification Experiment}

To evaluate the effectiveness of the proposed DCPV, we tested the proposed method in multiple public palmprint datasets. In the experiments, we utilized the same key across different IDs, emulating a stolen-token worst-case scenario~\cite{dong2022co}. 

The Equal Error Rate (EER) is a widely used metric to assess biometric verification performance. It denotes the value that the False Acceptance Rate (FAR) equals the False Rejection Rate (FRR). Thus, a lower EER indicates better performance, reflecting a more balanced and accurate biometric system. In our experiments, ``DCPV$\dagger$" and ``DCPV$\ddagger$" denote the proposed DCPV with only the first- and second-level protections, respectively. In contrast, ``Baseline" refers to the palmprint verification method employed without protective measures, serving as an approximate upper-bound performance. 

Our proposed method can deliver satisfactory results with minimal performance degradation across the IITD, PolyU, and NIR datasets. The limited performance degradation of DCPV$\ddagger$ highlights the effectiveness of using a negative database in palmprint verification. Additionally, DCPV exhibits commendable performance under Red, Green, and Blue lighting conditions. 
These results support that the proposed method successfully achieves satisfactory performance without the need to store the same cancelable template used for verification.


\begin{table}[!t]
    \centering
    \small
    \begin{threeparttable}
      \caption{EERs of the Proposed Method on Different Databases.}
\renewcommand\arraystretch{1.2}
        \begin{tabular}{cccccccc}
        \hline
 & Baseline& & DCPV$\dagger$ & & DCPV$\ddagger$ & & DCPV \\
         &  EER (\%)&  &  EER (\%)&  &  EER (\%) &  &  EER (\%)   \\  \hline
         IITD & 0.15448 & & 0.22924 &  & 0.47101 & & 0.36232   \\
         PolyU & 0.00044 &  & 0.00123 & & 0.00094 & & 0.00794 \\
         Red & 0 &  & 0 & & 0 & & 0   \\
         Green & 0 &  & 0 & & 0 & & 0  \\
         Blue & 0 &  & 0 & & 0 & & 0   \\
         NIR & 0 & & 0 & & 0.00005 & & 0.00007 \\ \hline
    \end{tabular}
    \begin{tablenotes}
        \item ``DCPV$\dagger$" and ``DCPV$\ddagger$" denote the proposed DCPV with only the first- and second-level protections, respectively.
        \end{tablenotes}
    \end{threeparttable}
    \label{tab:main_eer}
\end{table}

\begin{figure}
  \begin{minipage}[t]{0.49\linewidth}
  \centering
  \includegraphics[width=\textwidth]{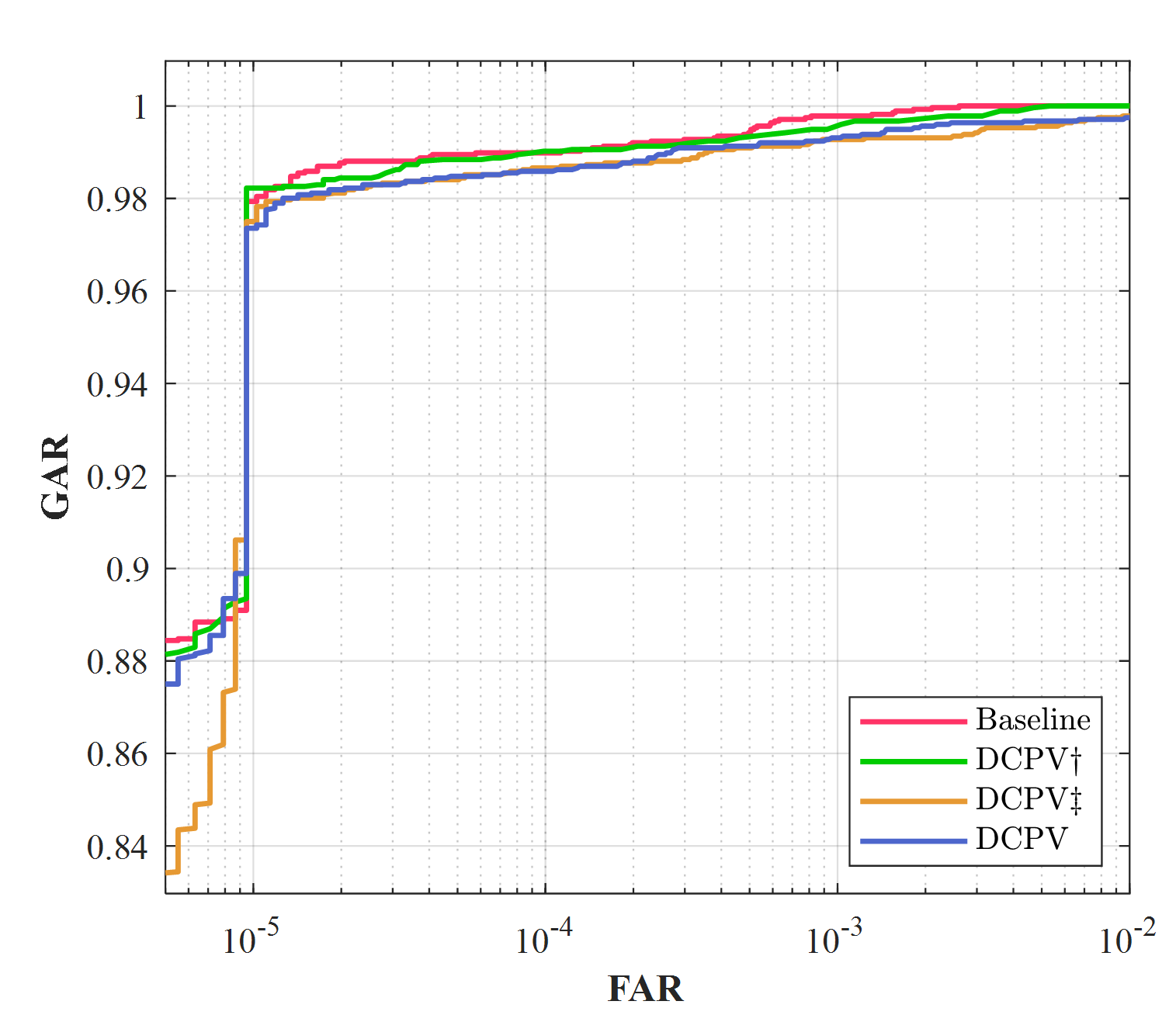}
   \centerline{(a)}
   \end{minipage}  
  \begin{minipage}[t]{0.49\linewidth}
  \centering
  \includegraphics[width=\textwidth]{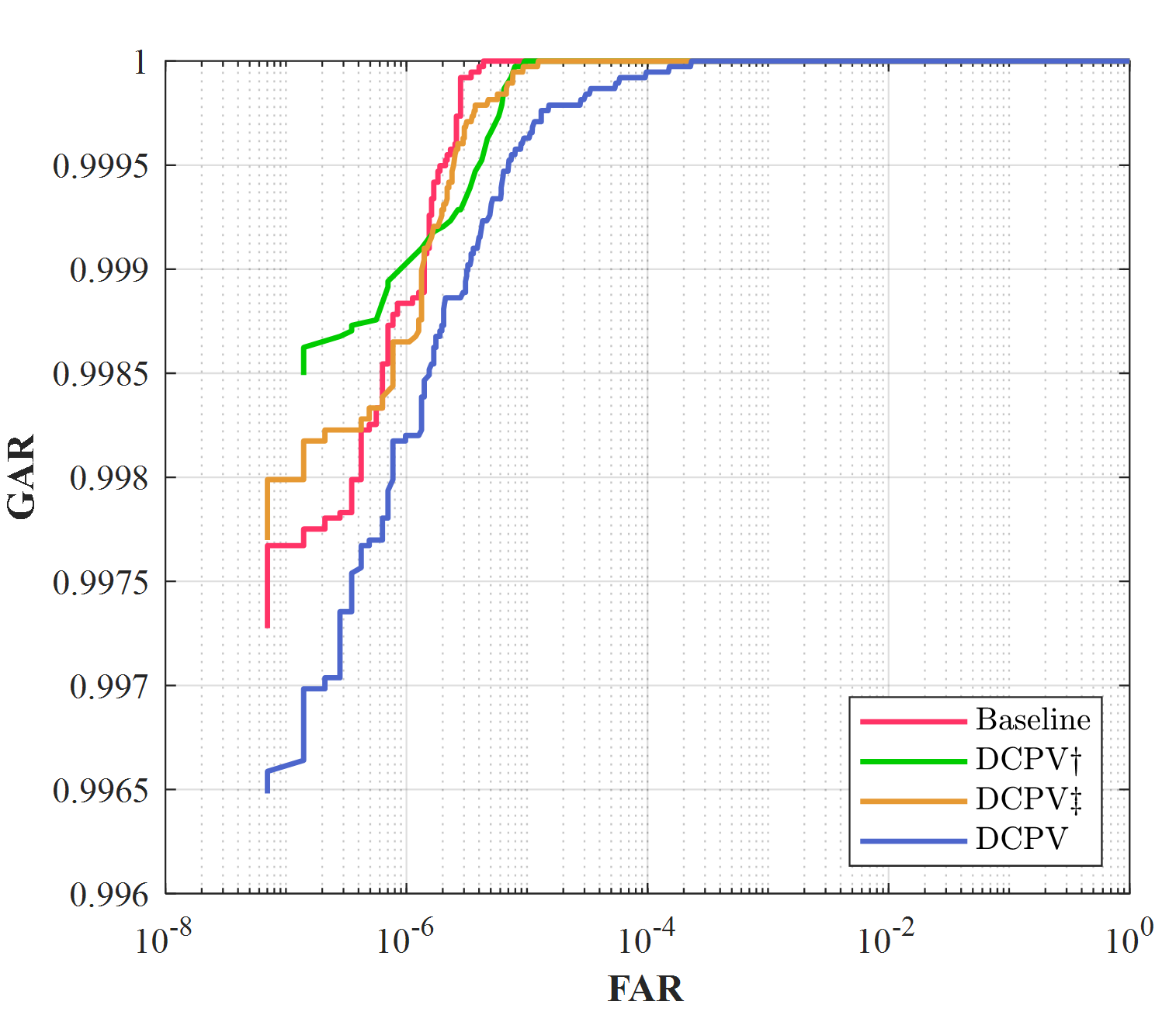}
   \centerline{(b)}
   \end{minipage}
     \caption{The ROC curves of the proposed method on different databases. (a)-(b) denotes the IITD, and PolyU results.}
  \label{fig:distri}
\end{figure}

The Genuine Acceptance Rate (GAR), defined as $GAR = 1 - FRR$, along with FAR, serves as another popular validation metric for verification. We compared the performance of the palmprint verification with different protection strategies using the Receiver Operating Characteristic (ROC) curve, which illustrates the relationship between GAR and FAR. This curve offers an in-depth evaluation of the balance between these two metrics. Optimal performance is achieved when the ROC curve is close to the top left corner of the plot, indicating a high GAR and low FAR. Due to the ability of DCPV to achieve minimal performance degradation across Red, Green, Blue, and NIR datasets, we only illustrate ROC curves for IITD and PolyU datasets. Therefore, this confirms the proposed method as an effective and competitive strategy for palmprint protection, balancing privacy safeguards and minimal verification performance degradation.

\begin{figure}
 \vspace{5pt}
  \begin{minipage}[t]{0.49\linewidth}
  \centering
  \includegraphics[width=\textwidth]{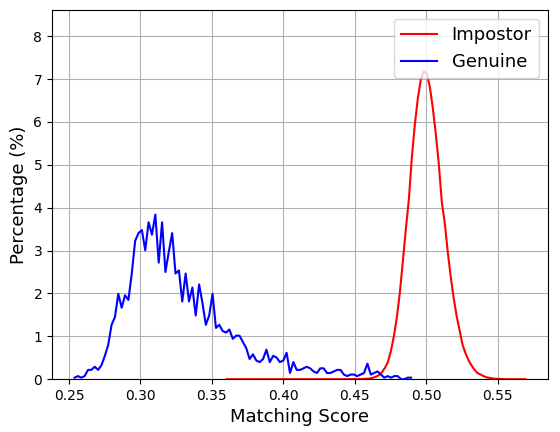}
   \centerline{(a)}
   \end{minipage}  
  \begin{minipage}[t]{0.49\linewidth}
  \centering
  \includegraphics[width=\textwidth]{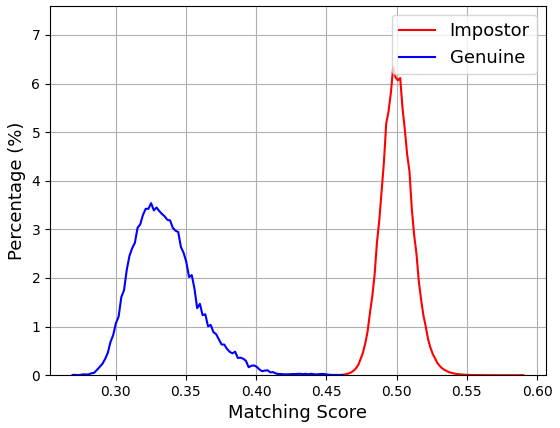}
   \centerline{(b)}
   \end{minipage} \\
     \begin{minipage}[t]{0.49\linewidth}
  \centering
  \includegraphics[width=\textwidth]{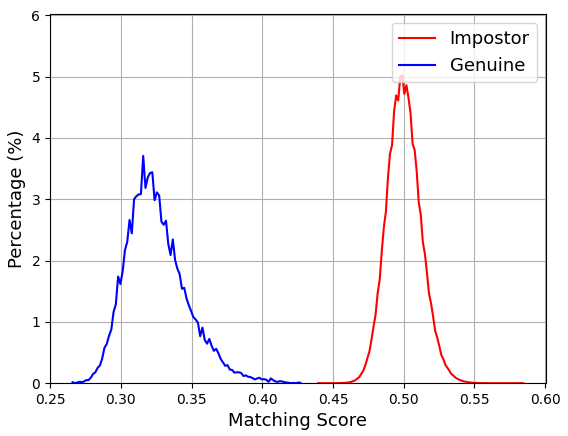}
   \centerline{(c)}
   \end{minipage}  
  \begin{minipage}[t]{0.49\linewidth}
  \centering
  \includegraphics[width=\textwidth]{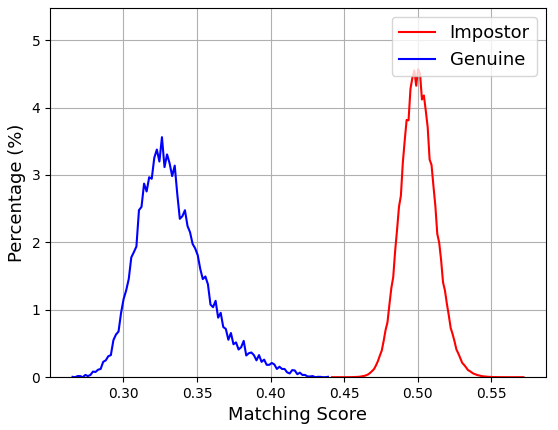}
   \centerline{(d)}
   \end{minipage} \\
     \begin{minipage}[t]{0.49\linewidth}
  \centering
  \includegraphics[width=\textwidth]{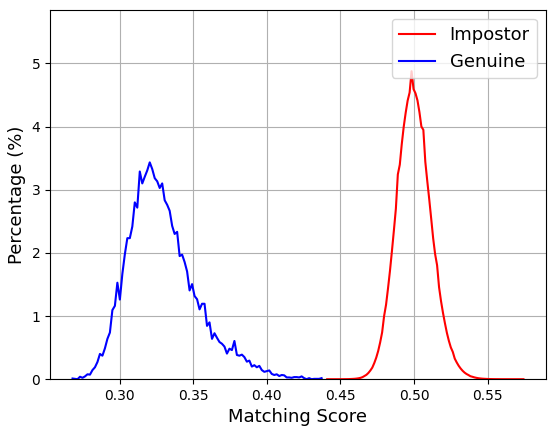}
   \centerline{(e)}
   \end{minipage}  
  \begin{minipage}[t]{0.49\linewidth}
  \centering
  \includegraphics[width=\textwidth]{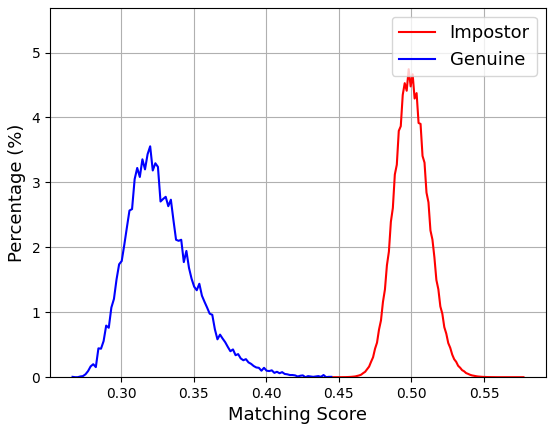}
   \centerline{(f)}
   \end{minipage}
     \caption{The matching distributions of the proposed method on different databases. (a)-(f) denotes the IITD, PolyU, Red, Green, Blue, and NIR results.}
  \label{fig:distribution}
\end{figure}

The genuine and imposter-matching distributions are shown in Fig.~\ref{fig:distribution}.
The results reveal substantial gaps between the imposter and genuine distributions. Notably, in the Red, Green, and Blue datasets, the two distributions do not overlap at all. This suggests that the proposed DCPV effectively maintains the network's discriminative ability. Additionally, to highlight the statistical discrepancy, we list the average and standard deviation (STD) of both distributions in different datasets in Tab.~\ref{tab:distribution}. The significant gap between the average genuine and imposter matching distances, along with the small variances in the distributions, is evident.

\begin{table}[!t]
    \centering
    \small
      \caption{The Statistical Indicators of Matching Distributions.}
      \renewcommand\arraystretch{1.2}
        \begin{tabular}{cccc}
        \hline
 & Genuine & & Imposter\\
         &   Avg $\pm$ STD& &   Avg $\pm$ STD\\  \hline
         IITD& $0.3290 \pm 0.0390$
 &  & $0.5002 \pm 0.0120$\\
         PolyU & $0.3355 \pm 0.0235$
 &    & $0.5002 \pm 0.0111$
\\
         Red& $0.3256 \pm 0.0218$
 &   & $0.5002 \pm 0.0120$
\\
         Green&   $0.3337 \pm 0.0241$
   & & $0.5003 \pm 0.0117$
\\
         Blue & $0.3295 \pm 0.0238$
    & & $0.5004 \pm 0.0117$
\\
         NIR&  $0.3280 \pm 0.0240$
  & & $0.5002 \pm 0.0120$
\\ \hline
    \end{tabular}
    \label{tab:distribution}
\end{table}



\noindent 

\subsection{Abaltaion Study}

\begin{table}[!t]
    \centering
    \small
        \begin{threeparttable}
      \caption{EERs(\%) of the Proposed Method based on CO$_3$Net.}
\renewcommand\arraystretch{1.2}
        \begin{tabular}{cccccccc}
        \hline
 & Baseline& & DCPV$\dagger$ & & DCPV$\ddagger$ & & DCPV \\ \hline
         IITD & 0.45270 & & 0.57971 &  & 0.72464 & & 0.68841   \\
         PolyU & 0.01609 &  & 0.02381 & & 0.01587 & & 0.02646 \\
         Red & 0 &  & 0 & & 0 & & 0   \\
         Green & 0.00021 &  & 0.00031 & & 0.00027 & & 0.00242  \\
         Blue & 0 &  & 0.00003 & & 0.00019 & & 0.00051   \\
         NIR & 0 & & 0 & & 0.00176 & & 0.00041 \\ \hline
    \end{tabular}
        \begin{tablenotes}
        \item ``DCPV$\dagger$" and ``DCPV$\ddagger$" denote the proposed DCPV with only the first- and second-level protection implemented.
        \end{tablenotes}
    \end{threeparttable}
    \label{tab:main_co3_eer}
    
\end{table}

\begin{figure}
  \vspace{5pt}
  \begin{minipage}[t]{0.488\linewidth}
  \centering
  \includegraphics[width=\textwidth]{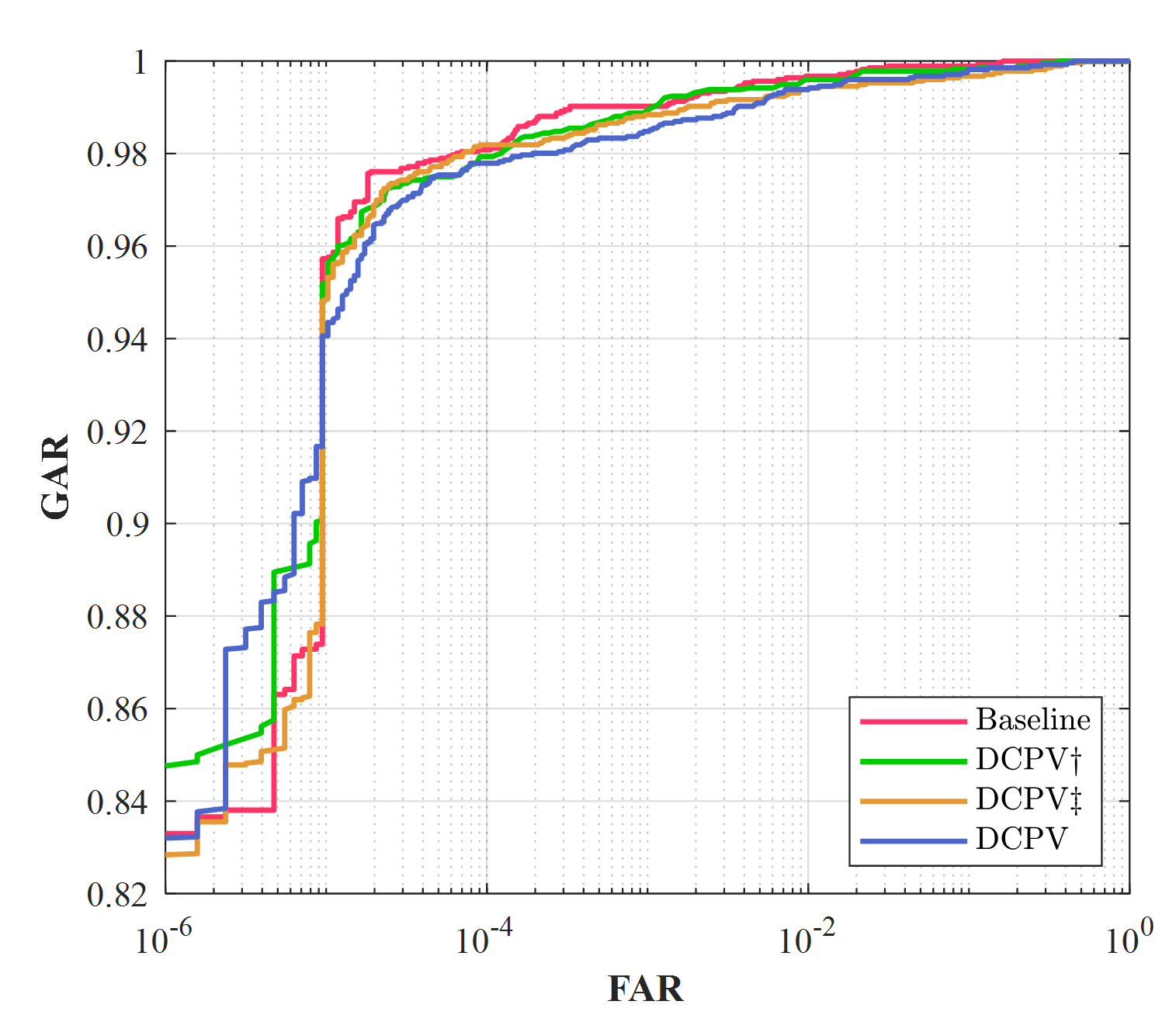}
   \centerline{(a)}
   \end{minipage}  
  \begin{minipage}[t]{0.492\linewidth}
  \centering
  \includegraphics[width=\textwidth]{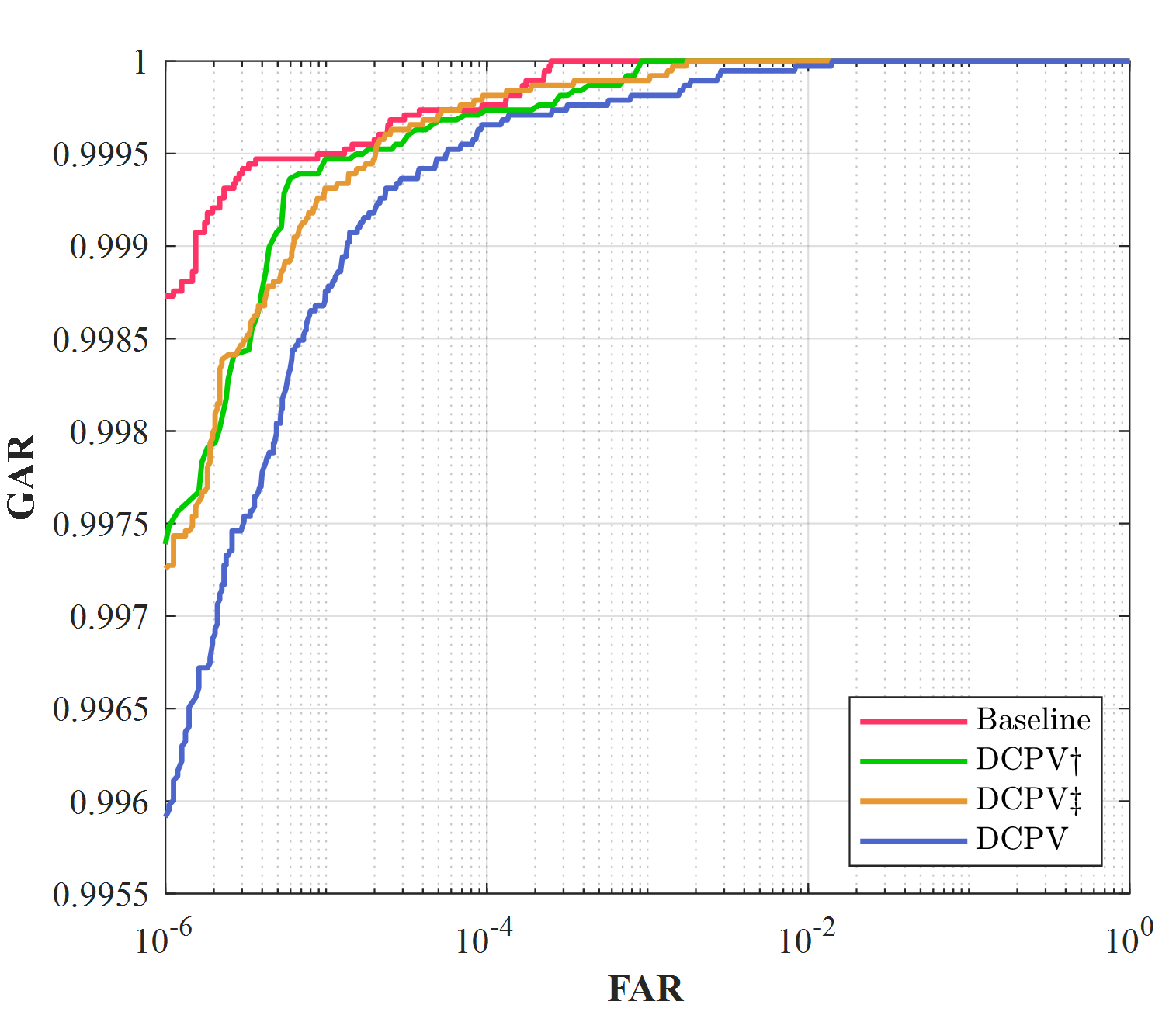}
   \centerline{(b)}
   \end{minipage}
     \caption{The ROC  of the proposed method on different databases based on CO3Net. (a)-(f) denotes the results on IITD and PolyU, respectively.}
  \label{fig:co3net}
\end{figure}

To validate the effectiveness and generalization of the proposed protection scheme, we further tested the performance of the proposed DCPV in another competition mechanism-based palmprint verification network CO$_3$Net~\cite{yang2023co3net}. The EERs on different datasets are illustrated in Tab.~\ref{tab:main_co3_eer}, and the ROC curves of CO$_3$Net on IITD and PolyU are shown in Fig.~\ref{fig:co3net}. The results illustrate that the proposed DCPV could maintain its performance with different palmprint feature extractors. In all datasets, the performance degradation compared with the Baseline is acceptable. This experiment demonstrates the effectiveness and generalization of the proposed method, highlighting its ability to integrate with various existing palmprint verification networks.

\begin{figure}[!t]
  \centering
  \includegraphics[width=.8\columnwidth]{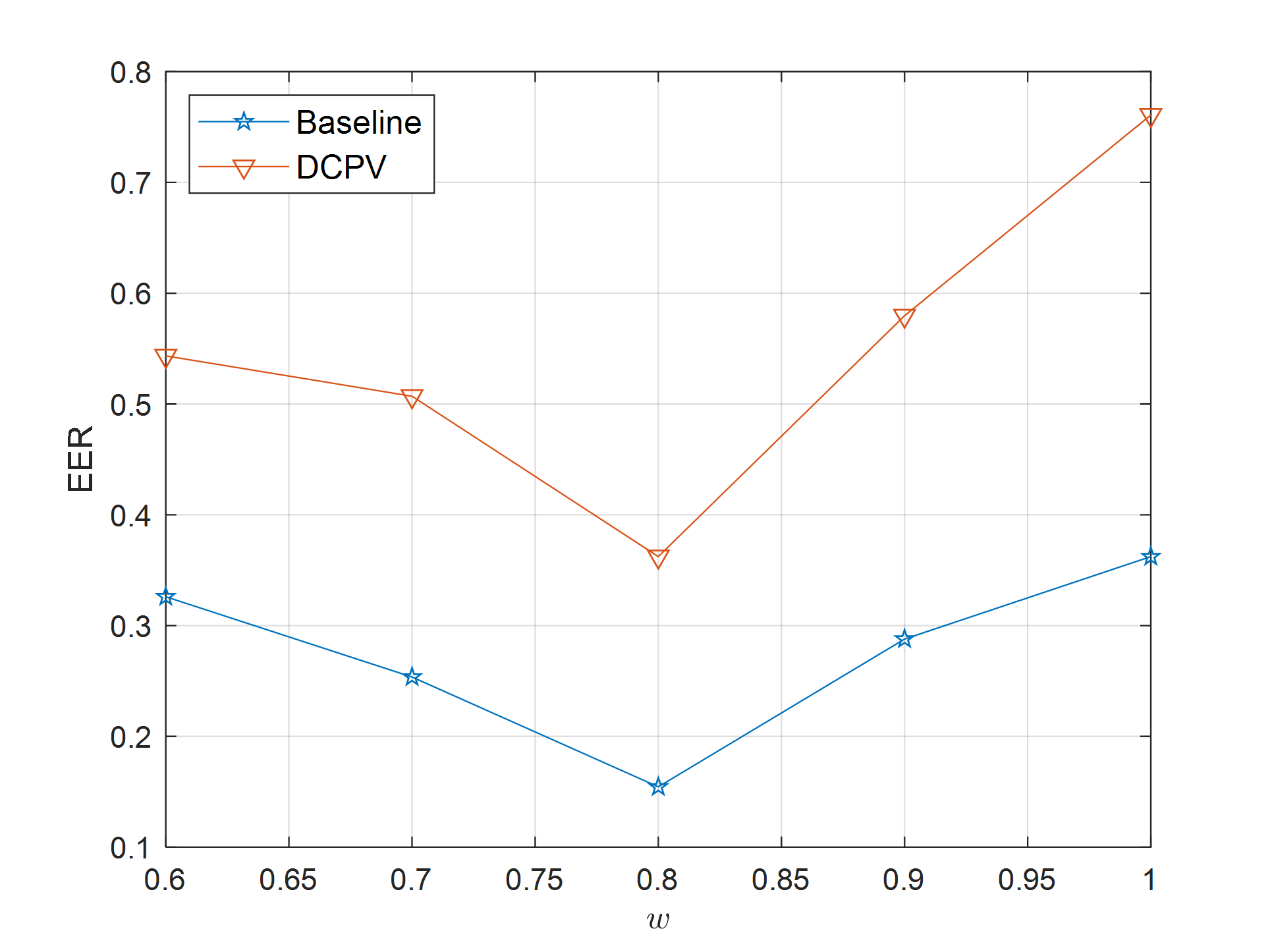}
     \caption{The ablation study about $w$ in Eq. (5).}
  \label{fig:ablation}
\end{figure}

Naive feature binarization may negatively impact performance. To mitigate this, we employ a contrastive loss in our approach to improve the discriminative power of binarized features, as shown in Eq.~\ref{eq:ltc}. We performed an ablation study on the weighting factor $w$ in Eq.~\ref{hy_loss}, with the findings presented in Fig.~\ref{fig:ablation}. Setting 
$w=1.0$ indicates that contrastive loss is not used during training. Our results demonstrate that incorporating contrastive loss significantly boosts the original verification performance. Furthermore, this loss notably reduces the performance difference between the Baseline and the DCPV model. Based on empirical experiments, we recommend setting 
$w=0.8$.

\begin{figure}
  \centering
  \begin{minipage}[t]{0.85\linewidth}
  \centering
  \includegraphics[width=\textwidth]{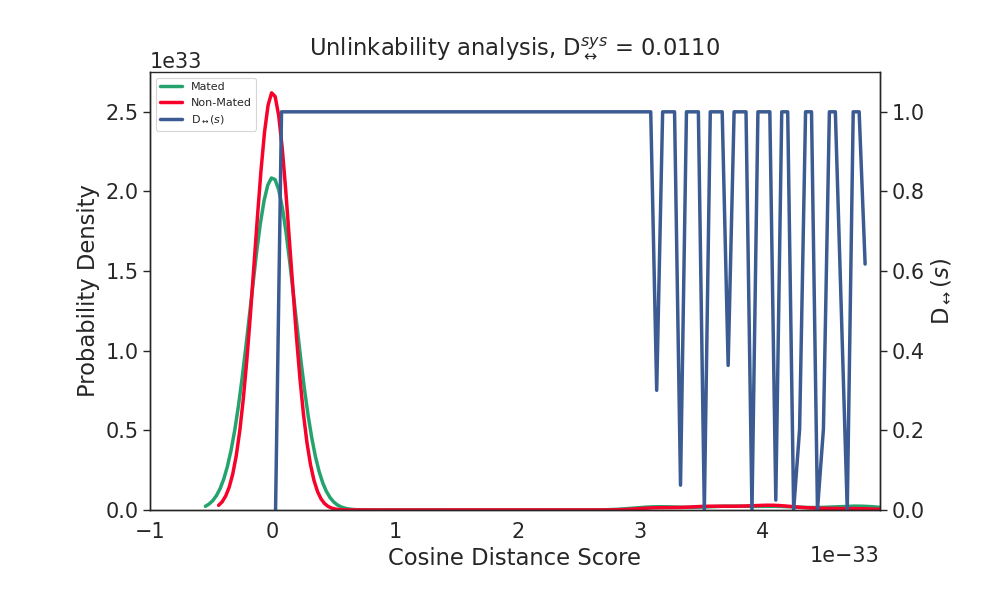}
   \centerline{(a)}
   \end{minipage}  \\
   \centering
  \begin{minipage}[t]{0.85\linewidth}
  \centering
  \includegraphics[width=\textwidth]{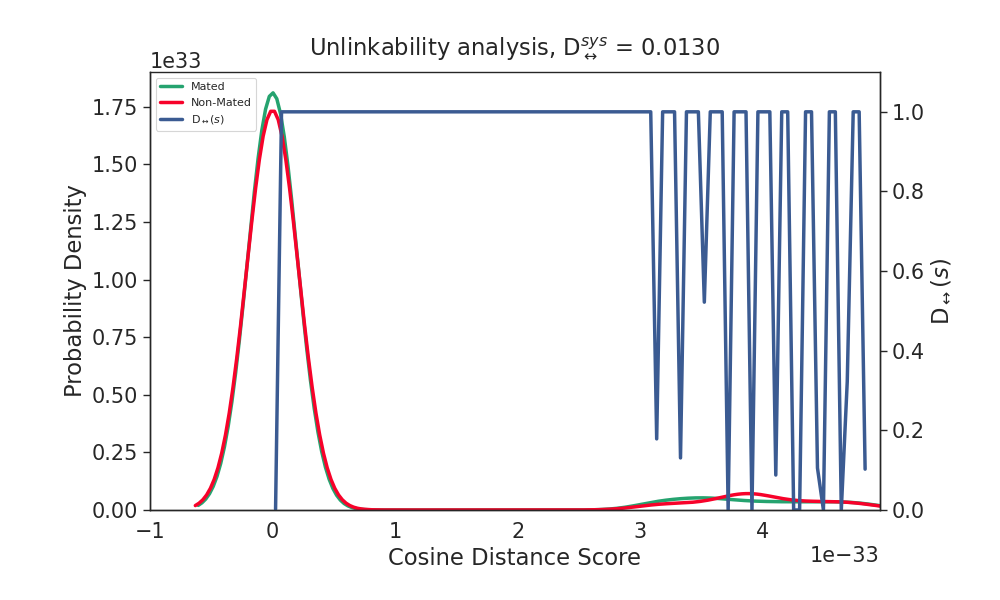}
   \centerline{(b)}
   \end{minipage} \\
   \centering
   \begin{minipage}[t]{0.85\linewidth}
  \centering
  \includegraphics[width=\textwidth]{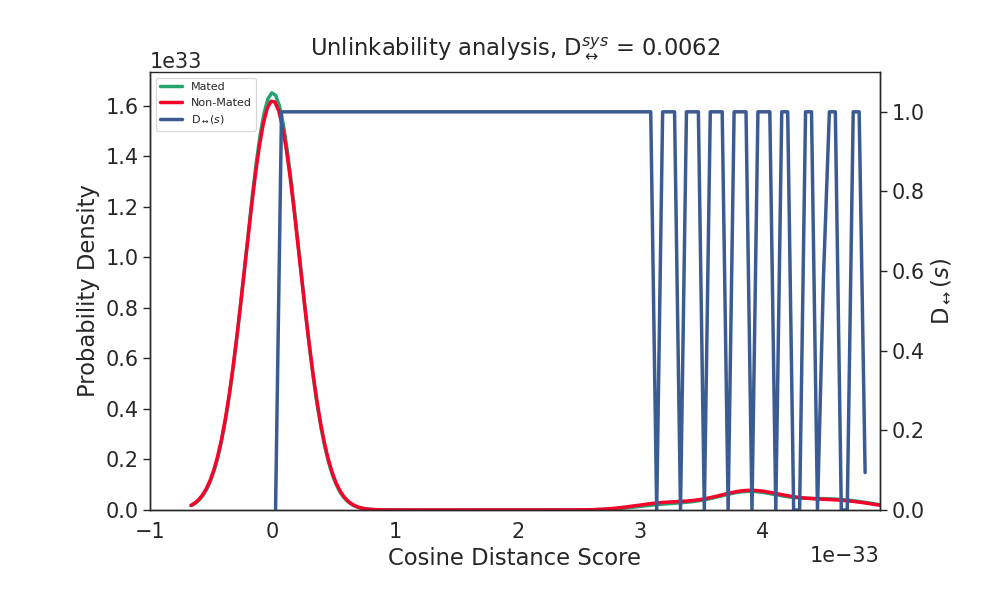}
   \centerline{(c)}
   \end{minipage}
     \caption{Unlinkability analysis of the proposed method. (a)-(c) represent DCPV$\dagger$, DCPV$\ddagger$ and DCPV, respectively.}
  \label{fig:unlink}
\end{figure}

\subsection{Unlinkability Analysis}

This subsection validates the unlinkability of the proposed DCPV using the protocol described in~\cite{gomez2017general}. The validation protocol relies on two distinct types of distributions: mated ($H_{m}$) and non-mated ($H_{nm}$) samples. Specifically, mated samples represent templates extracted from samples of the same ID but processed using different user-specific keys. Conversely, non-mated samples are templates extracted from samples of different IDs with different keys. To establish an unlinkable system, there must be significant overlap in the score distributions of mated and non-mated samples.

Then, two unlinkability measures could be calculated: the local measure and the global measure. 
Using these distributions, two unlinkability metrics are applied. \textit{i}) The local measure, denoted as $\mathrm{D}_{\leftrightarrow}(s)$, assesses the linkability of the system for specific linkage scores $s$, based on the likelihood ratio between score distributions. $\mathrm{D}_{\leftrightarrow}(s) \in[0,1]$ is defined across the entire score domain. A value of $\mathrm{D}_{\leftrightarrow}(s)=0$ signifies complete unlinkability, whereas $\mathrm{D}_{\leftrightarrow}(s)=1$ indicates full linkability between two transformed templates at score $s$. \textit{ii}) The global measure $D_{\leftrightarrow}^{sys} \in[0,1]$, independent of the score domain, offers an overall evaluation of the system's linkability. It is a fairer benchmark for comparing unlinkability across different systems. $D_{\leftrightarrow}^{sys}=1$ denotes complete linkability for all scores within the mated samples distribution, and $D_{\leftrightarrow}^{sys}=0$ indicates complete unlinkability across the entire score domain.

The IITD dataset was used to analyze both mated and non-mated sample distributions. These distributions were utilized to calculate the local measure $\mathrm{D}_{\leftrightarrow}(s)$, which then helps determine the global measure $D_{\leftrightarrow}^{sys}$, indicating the overall linkability of the system. Fig.~\ref{fig:unlink} illustrates the unlinkability curves for transformed templates created with DCPV$\dagger$, DCPV$\ddagger$, and DCPV, respectively. A noticeable overlap in these curves suggests minimal overall linkability for the proposed DCPV ($D_{\leftrightarrow}^{sys}=0.0062$). Consequently, the proposed system can be regarded as essentially unlinkable.

\section{Conclusions}
\label{sec:con}

This study introduces a novel dual-level cancelable palmprint verification framework, DCPV, explicitly designed for cloud verification scenarios. DCPV provides dual-level security. A key feature of DCPV is its ability to protect privacy and security against attacks on stored databases. Extensive experiments on several public datasets demonstrate the efficacy and robustness of DCPV in terms of both effectiveness and security. DCPV has shown commendable results in one-to-one matching scenarios, which constitute the verification task. However, it is not currently optimized for identification scenarios where the goal is to identify a subject from a large pool. Looking ahead, a key research direction from this work is to delve into and enhance identification scenarios. 

\bibliographystyle{IEEEtran}
\bibliography{ref}


 




\begin{IEEEbiography}[{\includegraphics[width=1in,height=1.25in,clip,keepaspectratio]{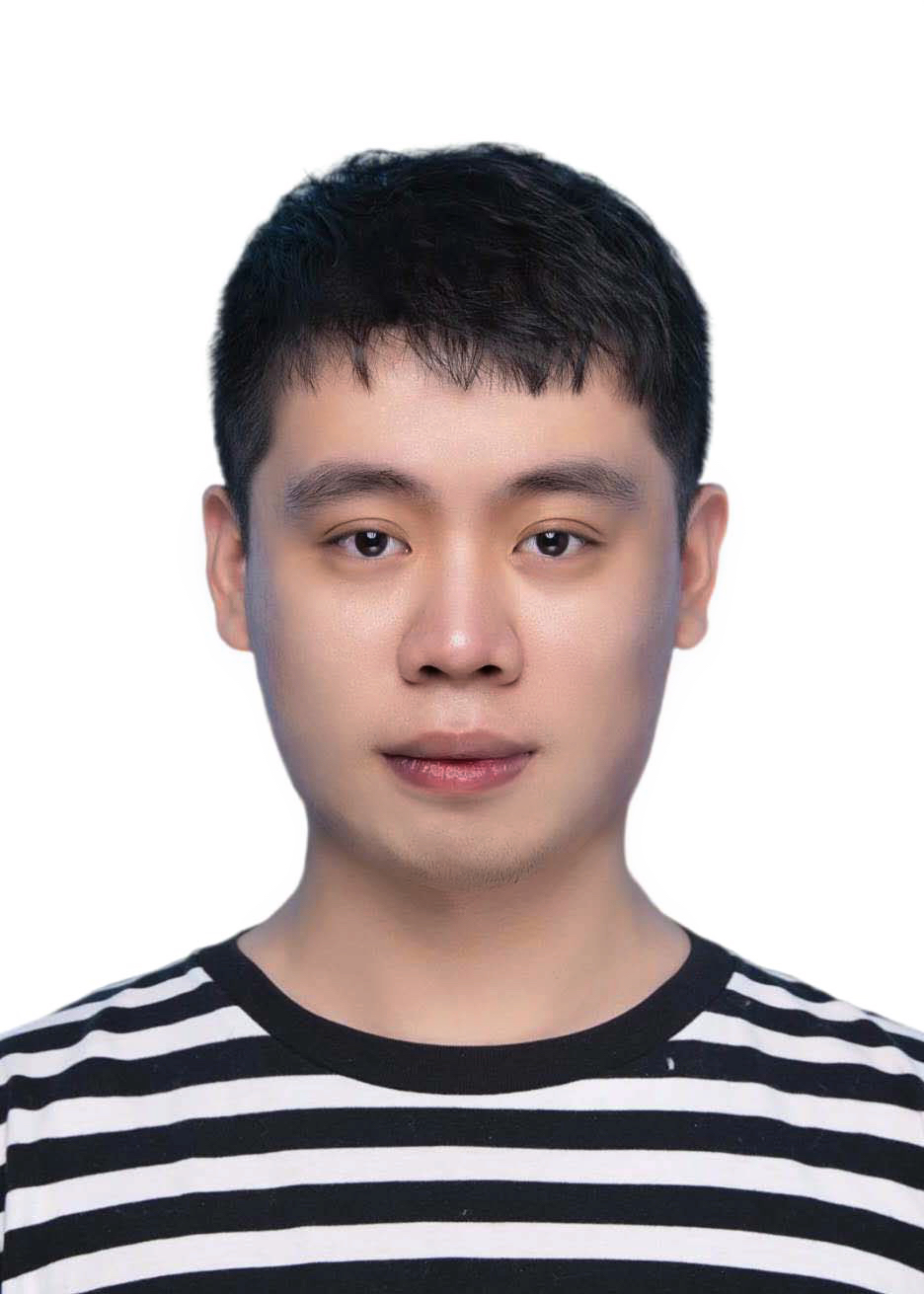}}]{Ziyuan Yang}
received the M.S. degree in computer science from the  School of Information Engineering, Nanchang University, Nanchang, China, in 2021. He is currently pursuing his Ph.D. degree with the College of Computer Science, Sichuan University, China. His research interests include biometrics, distributed learning, and security analysis.
\end{IEEEbiography}

\begin{IEEEbiography}[{\includegraphics[width=1in,height=1.25in,clip,keepaspectratio]{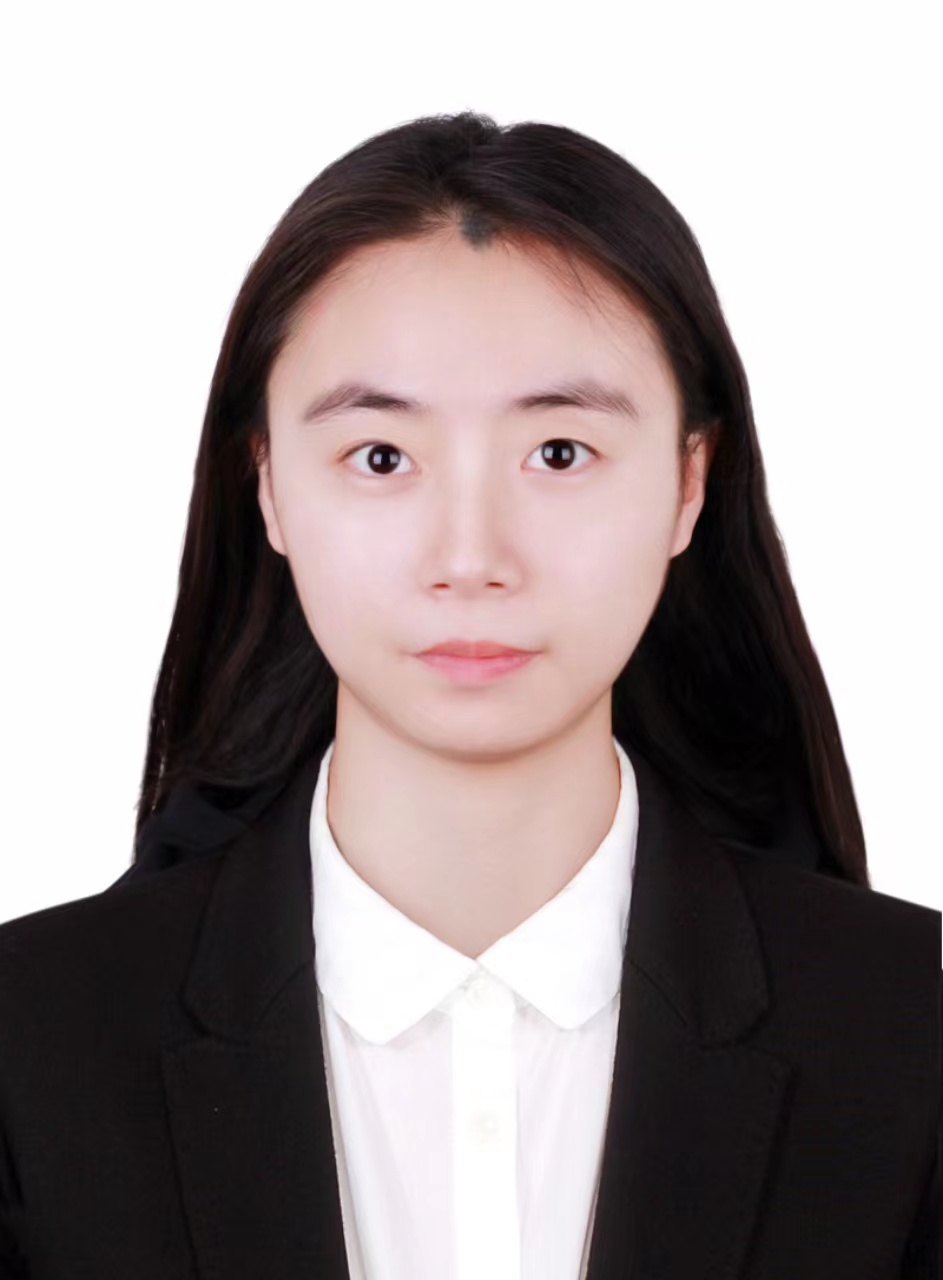}}]{Ming Kang}
received the M.E. degree from the School of Cyber Science and Engineering, Sichuan University, Sichuan, China, in 2023. She is currently pursuing her Ph.D. degree at the School of Cyber Science and Engineering, Sichuan University, China. Her research interests include deep learning and APT detection.
\end{IEEEbiography}

\begin{IEEEbiography}[{\includegraphics[width=1in,height=1.25in,clip,keepaspectratio]{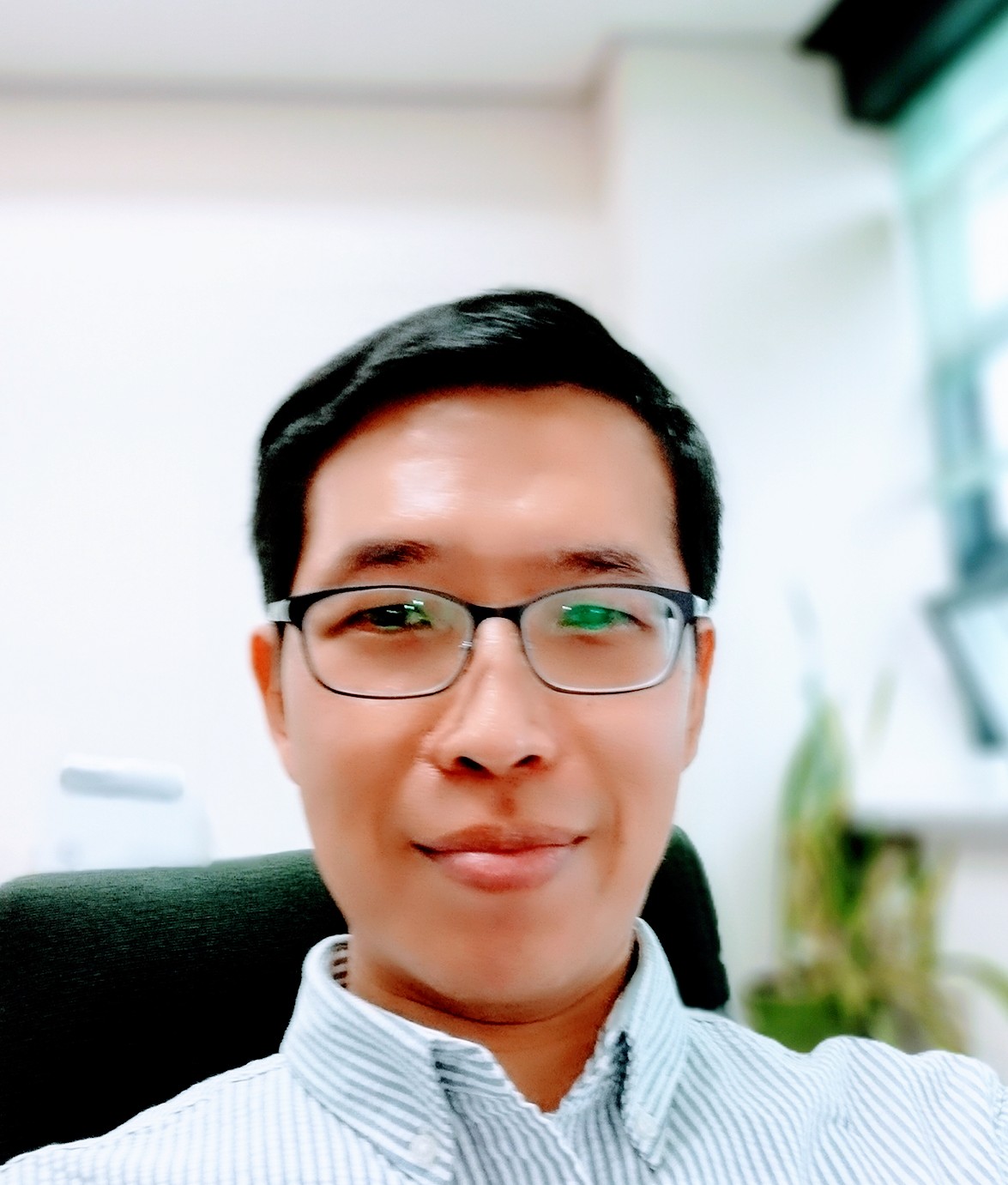}}]{Andrew Beng Jin Teoh}
(Senior Member, IEEE) obtained his BEng (Electronic) in 1999 and a Ph.D. degree in 2003 from the National University of Malaysia. He is currently a full professor in the Electrical and Electronic Engineering Department, College Engineering of Yonsei University, South Korea. His research for which he has received funding focuses on biometric applications and biometric security. His current research interests are Machine Learning and Information Security. He has published more than 350 international refereed journal papers, and conference articles, edited several book chapters, and edited book volumes. He served and is serving as a guest editor of the IEEE Signal Processing Magazine, and associate editor of IEEE TRANSACTIONS ON INFORMATION FORENSIC AND SECURITY, IEEE Biometrics Compendium and Machine Learning with Applications, Elsevier.
\end{IEEEbiography}

\begin{IEEEbiography}[{\includegraphics[width=1in,height=1.25in,clip,keepaspectratio]{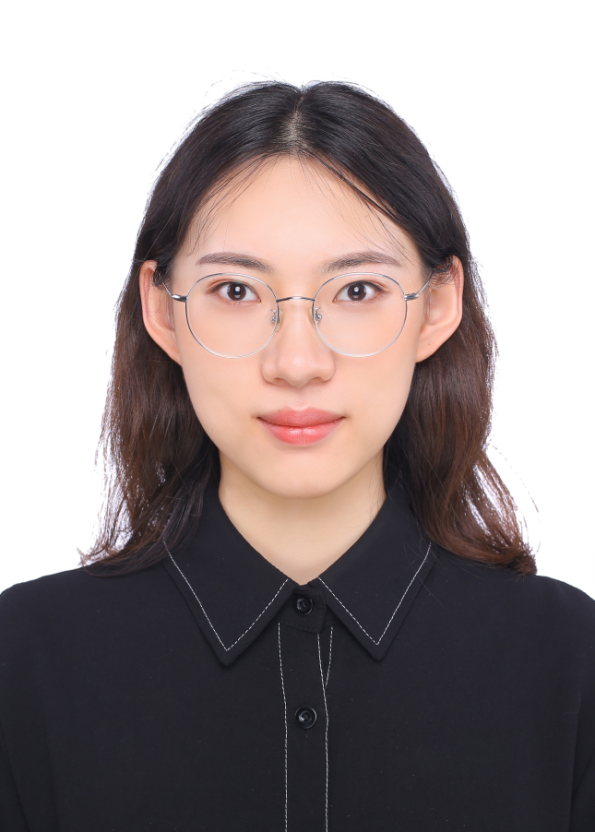}}]{Chengrui Gao}
received the M.S. degree from the School of Electronic Information, Sichuan University, Chengdu, China, in 2021. She is currently pursuing her Ph.D. degree at the College of Computer Science, Sichuan University. She is a visiting Ph.D. student at Yonsei University, South Korea. Her interests include biometrics and pattern recognition.
\end{IEEEbiography}

\begin{IEEEbiography}[{\includegraphics[width=1in,height=1.25in,clip,keepaspectratio]{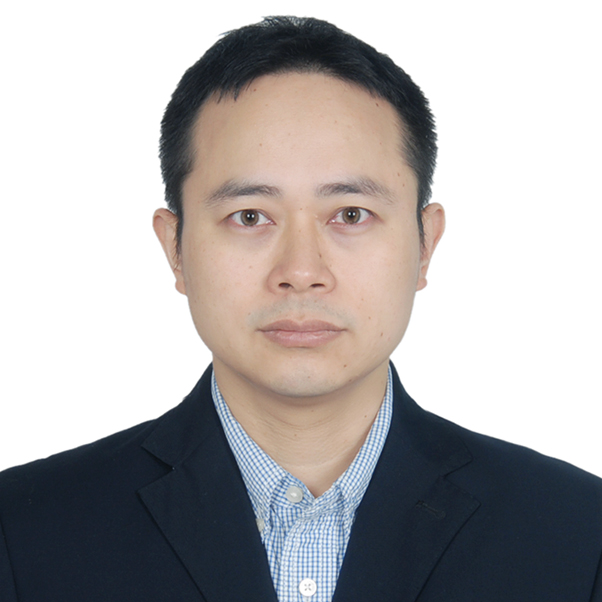}}]{Wen Chen}
received his M.S. degree in Software Engineering from Southwest University, Chongqing, China, and his Ph.D. degree in Computer Science from Sichuan University, Chengdu, China in 2011. He is currently an Associate Professor at Sichuan University. His research interests include artificial immune systems, adhoc networks, and information security. He worked as a reviewer for several Journals and Conferences. He is the recipient of the Award for Excellent Doctor Degree Dissertation of Sichuan Province.
\end{IEEEbiography}

\begin{IEEEbiography}[{\includegraphics[width=1in,height=1.25in,clip,keepaspectratio]{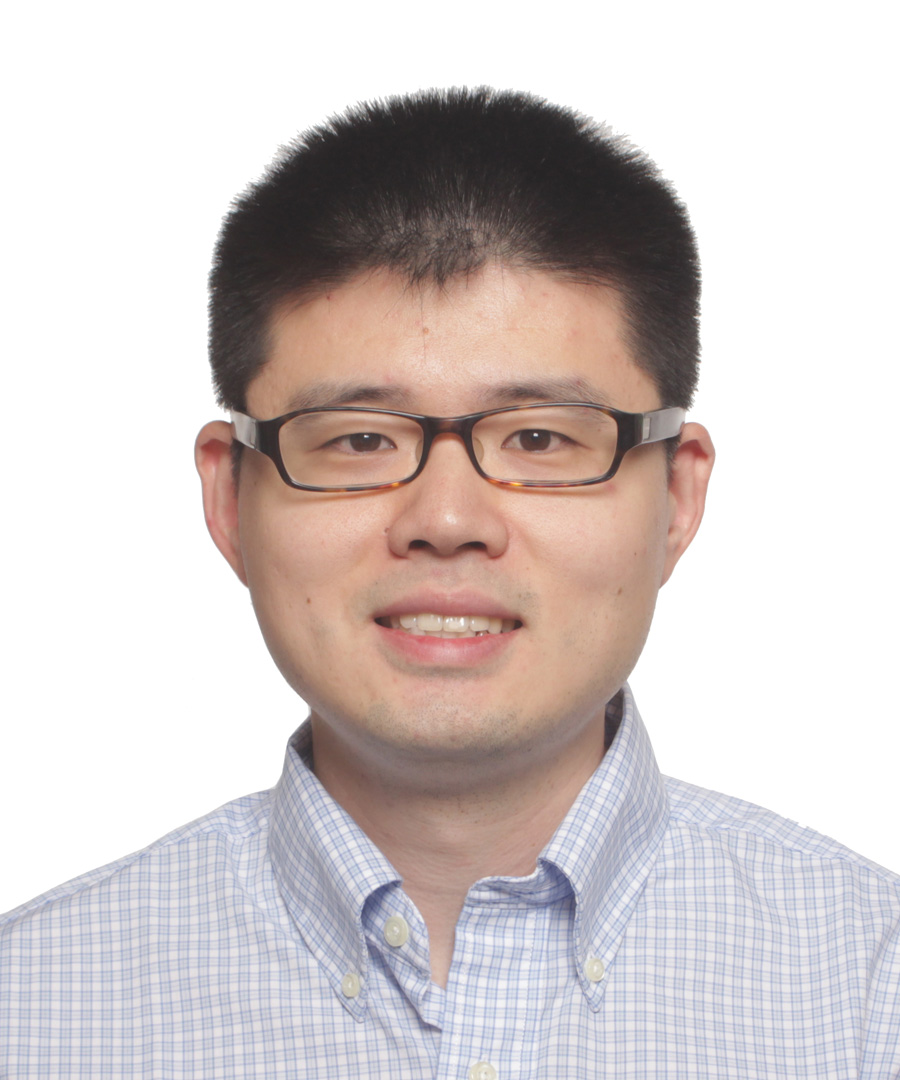}}]{Bob Zhang}
(Senior Member, IEEE) received the B.A. degree in computer science from York University, Toronto, ON, Canada, in 2006, the M.A.Sc. degree in information systems security from Concordia University, Montreal, QC, Canada, in 2007, and the Ph.D. degree in electrical and computer engineering from the University of Waterloo, Waterloo, ON, Canada, in 2011. After graduating from the University of Waterloo, he remained with the Center for Pattern Recognition and Machine Intelligence and later he was a Postdoctoral Researcher with the Department of Electrical and Computer Engineering, Carnegie Mellon University, Pittsburgh, PA, USA. He is currently an Associate Professor with the Department of Computer and Information Science, University of Macau, Macau. His research interests focus on biometrics, pattern recognition, and image processing. Dr. Zhang is a Technical Committee Member of the IEEE Systems, Man, and Cybernetics Society and an Associate Editor of IEEE TRANSACTIONS ON NEURAL NETWORKS AND LEARNING SYSTEMS, Artificial Intelligence Review, and IET Computer Vision.\end{IEEEbiography}

\begin{IEEEbiography}[{\includegraphics[width=1in,height=1.25in,clip,keepaspectratio]{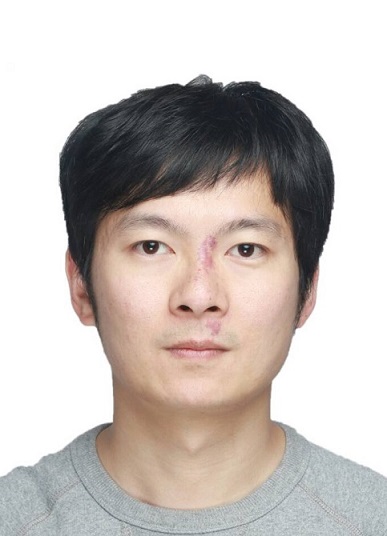}}]{Yi Zhang}
(Senior Member, IEEE) received the B.S., M.S., and Ph.D. degrees in computer science and technology from the College of Computer Science, Sichuan University, Chengdu, China, in 2005, 2008, and 2012, respectively. From 2014 to 2015, he was with the Department of Biomedical Engineering, Rensselaer Polytechnic Institute, Troy, NY, USA, as a Postdoctoral Researcher. He is currently a Full Professor with the School of Cyber Science and Engineering, Sichuan University, and is the Director of the deep imaging group (DIG). His research interests include medical imaging, compressive sensing, and deep learning. He authored more than 80 papers in the field of image processing. These papers were published in several leading journals, including IEEE TRANSACTIONS ON MEDICAL IMAGING, IEEE TRANSACTIONS ON COMPUTATIONAL IMAGING, Medical Image Analysis, European Radiology, Optics Express, etc., and reported by the Institute of Physics (IOP) and during the Lindau Nobel Laureate Meeting. He received major funding from the National Key R\&D Program of China, the National Natural Science Foundation of China, and the Science and Technology Support Project of Sichuan Province, China. He is a Guest Editor of the International Journal of Biomedical Imaging, Sensing and Imaging, and an Associate Editor of IEEE TRANSACTIONS ON MEDICAL IMAGING and IEEE TRANSACTIONS ON RADIATION AND PLASMA MEDICAL SCIENCES.\end{IEEEbiography}

\vfill

\end{document}